\documentclass[aps,prd,nofootinbib,showkeys,preprint,floatfix,showpacs]{revtex4}

\usepackage{amssymb}
\usepackage{amsmath}
\usepackage{amsfonts}
\usepackage{graphicx}
\usepackage{color}

\newcommand{\AddrAHEP}{
  {\it AHEP Group, Instituto de F\'{\i}sica Corpuscular --
    C.S.I.C./Universitat de Val{\`e}ncia \\
    Edificio de Institutos de Paterna, Apartado 22085,
  E--46071 Val{\`e}ncia, Spain}}

\newcommand{\AddrWur}{%
Institut f\"ur Theoretische Physik und Astronomie, 
Universit\"at W\"urzburg\\
Am Hubland, 
97074 Wuerzburg}

\def\gsim{\raise0.3ex\hbox{$\;>$\kern-0.75em\raise-1.1ex\hbox{$\sim\;$}}}
\def\lsim{\raise0.3ex\hbox{$\;<$\kern-0.75em\raise-1.1ex\hbox{$\sim\;$}}}
\newcommand{\nn}{\nonumber}

\newcommand{\Da}{\Delta m^2_{\textsc{A}}}

\begin{document}

\preprint{IFIC/10-52}  

\title{Supersymmetric mass spectra and the seesaw scale}

\author{M.~Hirsch} \email{mahirsch@ific.uv.es}\affiliation{\AddrAHEP}

\author{L.~Reichert} \email{reichert@ific.uv.es}\affiliation{\AddrAHEP}

\author{W. Porod} \email{porod@physik.uni-wuerzburg.de}\affiliation{\AddrWur}



\begin{abstract}
Supersymmetric mass spectra within two variants of the seesaw mechanism, 
commonly known as type-II and type-III seesaw, are calculated using full 
2-loop RGEs and minimal Supergravity boundary conditions. The type-II 
seesaw is realized using one pair of $15$ and $\overline{15}$ superfields, 
while the type-III is realized using three copies of $24_M$ superfields. 
Using published, estimated errors on SUSY mass observables attainable 
at the LHC  and in a combined LHC+ILC analysis, we calculate expected errors 
for the parameters of the models, most notably the seesaw scale. If SUSY 
particles are within the reach of the ILC, pure mSugra can be distinguished 
from mSugra plus type-II or type-III seesaw for nearly all relevant 
values of the seesaw scale. Even in the case when only the much less accurate 
LHC measurements are used, we find that indications for the seesaw can 
be found in favourable parts of the parameter space. Since our conclusions 
crucially depend on the reliability of the theoretically forecasted error 
bars, we discuss in some detail the accuracies which need to be achieved 
for the most important LHC and ILC observables before an analysis, such 
as the one presented here, can find any hints for type-II or type-III 
seesaw in SUSY spectra.

\end{abstract}

\maketitle


\section{Introduction}

In the minimal supersymmetric extension of the standard model (MSSM) 
all soft SUSY breaking mass terms are treated as free parameters, to 
be fixed at the electro-weak scale. However, these soft 
parameters potentially contain a wealth of information about physics 
at the high scale and understanding the nature of SUSY breaking will 
become the main challenge, if signals of SUSY are found at the LHC. 
Highly precise mass measurements will be needed to distinguish between 
different SUSY breaking schemes such as ``minimal supergravity'' 
(mSugra) \cite{Chamseddine:1982jx,Nilles:1983ge}, anomaly mediated 
SUSY breaking (AMSB) \cite{Giudice:1998xp,Chacko:1999am} or 
gauge mediated SUSY breaking (GMSB) \cite{Giudice:1998bp}, to 
name just the most familiar ones. 

However, all of the models mentioned above break SUSY at energies 
inaccessible for collider experiments. Thus, theoretical extrapolations 
from the TeV scale to the high energy scale will be needed and any 
``test'' of SUSY breaking schemes can at best take the form of a 
consistency check. Based on the results of
 \cite{AguilarSaavedra:2001rg,Weiglein:2004hn,AguilarSaavedra:2005pw} 
detailed calculations have been done, quantifying the accuracy with 
which such tests can be done using data from LHC and a possible ILC
\cite{Blair:2000gy,Blair:2002pg,Bechtle:2005vt,Lafaye:2007vs,Adam:2010uz}. 
However, these works concentrated on 
models with MSSM  particle content and thus did not attempt to take 
into account 
the observed non-zero neutrino masses. In this paper we study the 
prospects for the LHC and for a combined LHC+ILC analysis for finding 
indirect hints for the presence of a high-scale seesaw mechanism in SUSY 
spectra.

The MSSM assumes that R-parity is conserved and thus, just as in the 
standard model, neutrino masses vanish. Neutrino oscillation experiments 
\cite{Fukuda:1998mi,Ahmad:2002jz,Eguchi:2002dm,KamLAND2007}, however, have 
shown that at least two neutrino masses are non-zero \cite{Schwetz:2008er}. 
Among the myriad of possible models of neutrino masses, ``the seesaw'' 
mechanism \cite{Minkowski:1977sc,seesaw,MohSen} is undoubtedly the 
most popular one. The classical version of the seesaw 
\cite{Minkowski:1977sc,seesaw} introduces (at least two) fermionic singlets 
(``right-handed neutrinos'') with some large, but arbitrary Majorana 
mass $M_R$. The smallness of the observed neutrino masses is then a 
straightforward consequence of $m_{\nu}$ being inversely proportional 
to $M_R$. This variant of the seesaw is now usually called type-I 
seesaw. 

At tree-level there are only three realizations of the seesaw 
mechanism \cite{Ma:1998dn}. In addition to type-I, the seesaw can be 
generated by the exchange of a scalar triplet (type-II) 
\cite{Schechter:1980gr,Cheng:1980qt} or by a fermionic triplet in the 
adjoint representation (type-III) \cite{Foot:1988aq}. Common to all of 
them is that for $m_{\nu} \sim \sqrt{\Da} \sim 0.05$ eV, where 
$\Da$ is the atmospheric neutrino mass splitting, and couplings 
of order ${\cal O}(1)$ the scale of the seesaw is estimated to be 
very roughly $m_{SS} \sim 10^{15}$ GeV. 

Extending the standard model (SM) with a seesaw mechanism leaves no 
experimental signal apart from the neutrino masses themselves. The 
situation is different in the supersymmetric seesaw. There are two 
kind of measurements which potentially can give indirect information 
about the seesaw parameters: Lepton flavour violating (LFV) decays 
and superpartner mass measurements. 

The literature on LFV in SUSY seesaw is vast \cite{Hisano:1995cp}. 
It was pointed out already in \cite{Borzumati:1986qx} that LFV is 
practically unavoidable in SUSY seesaw, even if the SUSY breaking 
boundary conditions are completely flavour blind. However, for 
(3 generation) type-I and type-III seesaws the seesaw mechanism 
has more free parameters than there are observables in the neutral 
and charged lepton sectors. The three LFV entries in the left slepton 
mass matrix can then be made arbitrarily small (or large) independent 
from any neutrino physics and thus there are no definite predictions 
for LFV decays in SUSY seesaw. \footnote{The situation is different 
in ``minimal'' type-II seesaw. Here, {\em ratios} of different LFV 
decays are related to neutrino angles, if (a) mSugra 
boundary conditions are assumed \cite{Rossi:2002zb,Hirsch:2008gh} or (b) 
in schemes where the seesaw triplet is also responsible for SUSY breaking, 
see \cite{Joaquim:2006uz,Joaquim:2006mn,Brignole:2010nh}.}
Indeed, if any charged 
LFV is ever observed one would probably turn the argument around and try 
to learn indirectly about the unknown seesaw parameters instead
\cite{Hirsch:2008dy,Davidson:2001zk,Ibarra:2005qi}.

In contrast, there are only very few papers, which have studied the 
impact of the seesaw on SUSY particle masses.  Some aspects of type-I 
seesaw have been studied focusing on what can be learned from
precision measurements in the slepton sector
 \cite{Blair:2002pg,Freitas:2005et,Deppisch:2007xu,Kadota:2009sf}.
Moreover, in such a scenario a splitting between the masses of the
selectrons and smuons can occur which might be measurable at the LHC
\cite{Allanach:2008ib,Buras:2009sg,Abada:2010kj}.
  Changes 
in SUSY spectra can lead to changes in the expected relic density for 
the cold dark matter. The impact of large values of soft terms in the 
sneutrino sector \cite{Calibbi:2007bk} and of large values for the 
trilinear $A_0$ parameter \cite{Kadota:2009vq,Kadota:2009fg} have been 
studied in this context.

The relative scarcity of publications on SUSY spectra and the seesaw 
is probably explained by the fact that type-I seesaw, the undoubtedly 
most popular variant, adds only singlets to the MSSM particle content. 
If  the Yukawa couplings of these singlets are smaller than, say, 
the gauge couplings any effects of the right-handed neutrinos on the 
SUSY mass eigenvalues become negligibly small. This leaves only 
a rather small window for the seesaw scale, $m_{SS}$, say, roughly 
[$4\times 10^{14}, 1.2\times 10^{15}$] GeV where any measurable shifts 
in SUSY masses can be expected at all. And it is, of course, exactly 
this range for $m_{SS}$ where the largest values for LFV decays are 
expected.  Exceptions from this general rule can be found
in models where one departs from the universality
assumption of the mSUGRA parameters with huge soft SUSY breaking 
parameters in the seesaw sector one gets
larger effects \cite{Kang:2010zd}, 
in particular in the Higgs sector \cite{Heinemeyer:2010eg}.

Changes in SUSY spectra with respect to, say, mSugra expectations 
are expected to be much larger in type-II and type-III seesaws, but 
very little work has been done on these seesaw variants as well. 
Type-II and DM has been studied in \cite{Esteves:2009qr}, while for 
a study of the type-III seesaw with emphasis on spectra and LFV see 
\cite{Esteves:2010ff}. In \cite{Buckley:2006nv} it was pointed out, 
that one can form different combinations of soft SUSY breaking parameters, 
which at 1-loop order do not depend on the mSugra parameters. Consistent 
departures of these ``invariants'' from mSugra expectations could then 
be taken as indirect hints of the seesaw. (See, however, 
\cite{Hirsch:2008gh,Esteves:2010ff} for the importance of 2-loop effects 
on the ``invariants''.) 
The above papers \cite{Hirsch:2008gh,Esteves:2010ff,Buckley:2006nv} 
have pointed out, how type-II and type-III leave traces in 
SUSY spectra, in principle. They did not, however, attempt to 
{\em quantify the accuracies} needed to find hints of the seesaw 
in experimental mass measurements. To our knowledge
the current paper is
 the first step in this direction.

Here we calculate the low-energy SUSY spectra for type-II and type-III 
seesaw in mSugra and confront our theoretical results with expectations 
for the accuracy of SUSY mass measurements at the LHC and at a possible 
combined LHC+ILC analysis \cite{Weiglein:2004hn,AguilarSaavedra:2005pw}. 
Given the estimated errors on SUSY masses obtained in detailed simulations  
\cite{Weiglein:2004hn,AguilarSaavedra:2005pw} we calculate expected 
$\chi^2$-distributions for the two different seesaw models, in order 
to give a theoretical forecast on the expected errors on the model 
parameters, most notably the error on the ``determination'' of the seesaw 
scale $m_{SS}$. 

Our main results are the following: With the highly accurate mass 
measurements expected for the ILC it should be possible to distinguish 
between pure mSugra (i.e. a model with no seesaw at all) and mSugra 
plus type-II or type-III seesaw for  almost {\em any relevant 
value of $m_{SS}$}, 
if at least some SUSY particles are within kinematical reach of the ILC. 
We find it noteworthy, however, that even with the much less accurate 
data, expected from LHC measurements only, it seems possible to distinguish 
pure mSugra from the mSugra plus seesaw models in some favorable parts 
of parameter space. Obviously, all our results depend crucially on the 
- currently only theoretically estimated - errors, with which SUSY mass 
can be measured at LHC and ILC. We therefore discuss in some details 
what are the observables needed and the required error bars on these 
observables, before an analysis, such as the one presented here, can 
find any hints of the seesaw in SUSY spectra. 

The rest of this paper is organized as follows. In the next section 
we summarize the three different variants of the seesaw, to 
set up the notation. We embed the new particles
required by the different seesaw mechanisms in complete $SU(5)$ multiplets
in order to maintain the successful unification of 
gauge couplings observed in the MSSM. Section 3 contains the 
bulk of this paper. It first defines our setup, lists the 
observables we use and then discusses SUSY spectra in the 
different models. With these results we then proceed to calculate 
theoretical $\chi^2$ distributions. We first discuss a combined 
LHC+ILC analysis and then go to the case of using only the less accurate 
LHC data. Our results, of course, depend crucially on the accuracy 
with which the SUSY masses can be measured in future accelerator 
experiments. We therefore dedicate a subsection to discuss in 
detail the requirements for the accuracies on the most important 
observables needed for our analysis. We then close with a short 
discussion and outlook.

\section{Supersymmetric seesaws}

In this section we briefly recall the main features of the three
tree-level variants of the seesaw.  A more detailed discussion including 
the embedding in $SU(5)$ can be found in \cite{Borzumati:2009hu}. 
For brevity, we will discuss only the superpotential terms.
In all cases, we start with the MSSM superpotential
\begin{eqnarray}\label{eq:WMSSM}
W_{MSSM}& = & {\widehat U}^c Y_u {\widehat Q} \cdot {\widehat H}_u
         - {\widehat D}^c Y_d {\widehat Q} \cdot {\widehat H}_d
         - {\widehat E}^c Y_e {\widehat L} \cdot {\widehat H}_d 
              + \mu {\widehat H}_u \cdot {\widehat H}_d \thickspace . 
\end{eqnarray}
Here, $A \cdot B = A_1 B_2 - A_2 B_1$ denotes the $SU(2)$ invariant 
product of two $SU(2)$ doublets and $Y_{\alpha}$ are ($3,3$) Yukawa 
coupling matrices, as usual. Only for completeness we mention that a 
seesaw type-I is obtained by adding to the above superpotential:
\begin{eqnarray}\label{eq:WtypeI}
 W_{I}& = &  {\widehat N}^c Y_\nu {\widehat L} \cdot {\widehat H}_u
          + \frac{1}{2}{\widehat N}^c M_R  {\widehat N}^c \thickspace .
\end{eqnarray}
Integrating out the heavy singlets, after electro-weak symmetry breaking, 
eq. (\ref{eq:WtypeI}) leads to the famous seesaw formula
\begin{equation}
m_\nu = - \frac{v^2_u}{2} Y^T_\nu M^{-1}_R Y_\nu.
\label{eq:mnuI}
\end{equation}
The atmospheric neutrino mass scale requires at least one neutrino 
to be heavier than $\sqrt{\Da} \sim 0.05$ eV. For Yukawas of 
order ${\cal O}(1)$ this results in a ``seesaw scale'' of roughly 
$M_R \simeq 10^{15}$ GeV. Since eq. (\ref{eq:WtypeI}) only adds 
singlets to eq. (\ref{eq:WMSSM}) one does in general not expect any 
sizeable changes in the supersymmetric mass spectra at the electro-weak 
scale, see also the discussion in the introduction and in 
\cite{Esteves:2010ff}. In section \ref{subsec:spec} we 
will discuss, how the introduction of non-singlets changes the expected 
SUSY spectra and comment on the expected change for type-I.

In supersymmetric models the simplest way to generate a type-II, 
while maintaining gauge coupling unification, is to add a pair 
of $15$-plets of $SU(5)$ to eq.  (\ref{eq:WMSSM}). The $SU(5)$ 
invariant superpotential than reads 
\begin{eqnarray}\label{eq:pot15}
W & = & \frac{1}{\sqrt{2}}{\bf Y}_{15} {\bar 5}_M \cdot 15 \cdot {\bar 5}_M 
   + \frac{1}{\sqrt{2}}\lambda_1 {\bar 5}_H \cdot 15 \cdot {\bar 5}_H 
+ \frac{1}{\sqrt{2}}\lambda_2 5_H \cdot \overline{15} \cdot 5_H 
+ {\bf Y}_5 10 \cdot {\bar 5} \cdot {\bar 5}_H \nonumber \\ 
 & + & {\bf Y}_{10} 10_M \cdot 10_M \cdot 5_H + M_{15} 15 \cdot \overline{15} 
+ M_5 {\bar 5}_H \cdot 5_H.
\end{eqnarray}
Here, ${\bar 5}_M$ and $10_M$ are the usual $SU(5)$ matter multiplets and 
${5}_H =(H^c,H_u)$ and ${\bar 5}_H=({\bar H}^c,H_d)$. Under 
$SU(3)\times SU(2) \times U(1)$ the $15$-plet decomposes as  
\cite{Rossi:2002zb}
\begin{eqnarray}\label{eq:15}
15 & = &  S + T + Z  \thickspace,\\ \nonumber
S & \sim  & (6,1,-\frac{2}{3}), \hskip10mm
T \sim (1,3,1), \hskip10mm
Z \sim (3,2,\frac{1}{6}).
\end{eqnarray}
Below the GUT scale, $M_{G}$, in the $SU(5)$-broken 
phase the superpotential reads
\begin{eqnarray}\label{eq:broken}
W_{II} & = &  \frac{1}{\sqrt{2}}(Y_T \widehat{L} \widehat{T}_1  \widehat{L} 
+  Y_S \widehat{D}^c \widehat{S}_1 \widehat{D}^c) 
+ Y_Z \widehat{D}^c \widehat{Z}_1 \widehat{L}   \nonumber \\
& + & \frac{1}{\sqrt{2}}(\lambda_1 {\widehat H}_d \widehat{T}_1 {\widehat H}_d 
+\lambda_2 {\widehat H}_u \widehat{T}_2 {\widehat H}_u) 
+ M_T \widehat{T}_1 \widehat{T}_2 
+ M_Z \widehat{Z}_1 \widehat{Z}_2 + M_S \widehat{S}_1 \widehat{S}_2,
\end{eqnarray}
where fields with index 1 (2) originate from the $15$-plet 
($\overline{15}$-plet). The first term in eq.~(\ref{eq:broken}) 
is responsible for the generation of neutrino masses, which 
at low energies are given by 
\begin{eqnarray}\label{eq:ssII}
m_\nu=\frac{v_u^2}{2} \frac{\lambda_2}{M_T}Y_T.
\end{eqnarray}
Similar to type-I, the seesaw scale is estimated to be $\frac{M_T}
{\lambda_2} \simeq 10^{15} {\rm GeV}$. 

In the case of a seesaw model type-III one needs new fermions $\Sigma$ 
at the high scale belonging to the adjoint representation of $SU(2)$. 
The simplest complete $SU(5)$ embedding possible is the 24-plet 
\cite{Ma:1998dn}. The superpotential of the unbroken $SU(5)$ is then 
\begin{eqnarray}\label{eq:spot5}
W & = & \sqrt{2} \, {\bar 5}_M {\bf Y_5} 10_M {\bar 5}_H 
          - \frac{1}{4} 10_M {\bf Y_{10}} 10_M 5_H 
  +  5_H 24_M Y^{III}_N{\bar 5}_M +\frac{1}{2} 24_M M_{24}24_M \thickspace.
\end{eqnarray}
The $24_M$ decomposes under  $SU(3)\times SU(2) \times U(1)$ as 
\begin{eqnarray}\label{eq:def24}
24_M & = &(1,1,0) + (8,1,0) + (1,3,0) + (3,2,-5/6) + (3^*,2,5/6) \thickspace, 
\\ \nn
   & = & \widehat{B}_M + \widehat{G}_M + \widehat{W}_M + \widehat{X}_M + 
\widehat{\bar X}_M \thickspace.
\end{eqnarray}
The $\widehat{B}_M$ has the same quantum numbers as the ${\widehat N}^c$, 
while the fermionic component of the $\widehat{W}_M$ corresponds to 
the $\Sigma$. Thus, the $24_M$ always produces a combination of the 
type-I and type-III seesaw. 

In the $SU(5)$ broken phase the superpotential contains 
\begin{eqnarray}\label{eq:spotIII}
 W_{III} & = &    \widehat{H}_u( \widehat{W}_M Y_W - \sqrt{\frac{3}{10}} 
               \widehat{B}_M Y_B) \widehat{L}
 + \widehat{H}_u \widehat{\bar X}_M Y_X \widehat{D}^c \nonumber \\
         & & + \frac{1}{2} \widehat{B}_M M_{B} \widehat{B}_M 
         + \frac{1}{2}\widehat{G}_M M_{G} \widehat{G}_M 
          + \frac{1}{2} \widehat{W}_M M_{W} \widehat{W}_M 
          + \widehat{X}_M M_{X} \widehat{\bar X}_M . 
\end{eqnarray}
Integrating out the heavy fields, as before, leads to
\begin{equation}
m_\nu = - \frac{v^2_u}{2} \left( \frac{3}{10} Y^T_B M^{-1}_B Y_B 
+ \frac{1}{2} Y^T_W M^{-1}_W Y_W \right). 
\label{eq:mnu_seesawIII}
\end{equation}
There are two contributions: (i) from the gauge singlet and (ii) from 
the $SU(2)$ triplet. Starting with a common $Y^{III}_N$, $Y^T_B$ 
evolves slightly differently from $Y^T_W$ under the RGEs. Thus, 
in principle two non-zero neutrino masses are generated from one 
$24_M$ only. However, the ratio of the two non-zero neutrino masses 
generated in the RGE running is much too tiny to explain the observed 
neutrino data and thus at least 2 copies of $24_M$ are needed for a 
realistic neutrino mass spectrum. 
In our numerical calculations we use 3 copies of $24_M$, motivated 
by the observed 3 generations.With $\forall Y_B^{ij}/Y_W^{ij} \simeq 1$, 
one can simplify eq. 
(\ref{eq:mnu_seesawIII}) to
\begin{equation}
m_\nu = - v^2_u  \frac{4}{10} Y^T_W M^{-1}_W Y_W 
\label{eq:mnu_seesawIIIa}
\end{equation}
The scale of $M_W$ is then estimated to be $m_{SS}\sim 8\times 10^{14}$ GeV
for $Y_W^{ij}= O(1)$.

\section{{\label{sec:results}Results and discussion}}

In subsection \ref{subsec:setup} we will define our setup 
and discuss the input observables. In \ref{subsec:spec} we 
discuss the SUSY spectra in the different models and how 
the observables change under changes of the seesaw scale. 
Section \ref{subsec:LHCandILC} we present our results for a 
combined LHC+ILC analysis, while \ref{subsec:LHC} shows the 
results for an analysis using only LHC data. 
Section \ref{subsec:errors} discusses the accuracies on the 
different observables, which need to be achieved experimentally, 
before any conclusions on the presence (or absence) of a  
seesaw mechanism can be drawn. Given the inherent unreliability of 
theoretical error forecasts in general, section \ref{subsec:errors} 
can be considered to contain the central parts of the 
current paper.

\subsection{{\label{subsec:setup}Setup, observables and data input}}

All the numerical results shown in the following have been obtained 
with the programme package SPheno \cite{Porod:2003um,SPheno}. The 
RGE equations, complete at the 2-loop order, have been calculated 
and incorporated into SPheno with the help of SARAH 
\cite{Staub:2008uz,Staub:2009bi,Staub:2010jh}. Details and discussion 
of the implementation can be found in \cite{Esteves:2010ff}.

To completely specify the low-energy SUSY spectra, we have to assume a
specific SUSY breaking scenario. In this paper we use mSugra. mSugra
is defined at the GUT-scale, $M_G$, by: a common gaugino mass
$M_{1/2}$, a common scalar mass $m_0$ and the trilinear coupling
$A_0$, which gets multiplied by the corresponding Yukawa couplings to
obtain the trilinear couplings in the soft SUSY breaking
Lagrangian. In addition, at the electro-weak scale, $\tan\beta
=v_u/v_d$ is fixed. Here, as usual, $v_d$ and $v_u$ are the vacuum
expectation values (vevs) of the neutral component of $H_d$ and $H_u$,
respectively.  Finally, the sign of the $\mu$ parameter has to be
chosen.

In the following we will call ``pure mSugra'', pmSugra for short, 
the version of the model with no seesaw mechanism at all. Note 
that this is equivalent to putting the seesaw scale $m_{SS}$ 
equal to the GUT scale $M_G$. For brevity, we will call ``tpye-II'' 
and ``type-III'' models with mSugra boundary conditions, to which 
on top of the MSSM particle content the corresponding ``seesaw 
particles'', as specified in the previous section, are added at 
scale $m_{SS}$. 

In our numerical calculations we concentrate on some selected sets 
of mSugra parameters. This is mainly motivated by the exorbitant 
amount of CPU time a full scan over the mSugra space would require, 
see also the discussion below. The points we have studied are 
SPS1a' \cite{AguilarSaavedra:2005pw} and the points SPS1b and 
SPS3 \cite{Allanach:2002nj}. In addition, for reasons explained in 
section \ref{subsec:spec}, we consider a few more points with 
modified mSugra parameters. We will call these points MSP-1 
($m_0,M_{1/2}, \tan\beta,A_0$) = ($70,400,10,-300$), MSP-2 
($220,700,30,0$) and MSP-3 ($120,720,10,0$). MSP-1 is similar 
to SPS1a' but with a larger value of $M_{1/2}$, MSP-2 is a 
point with larger $\tan\beta$ and MSP-3 is similar to SPS3, but 
again with a larger value of $M_{1/2}$. All these points 
choose $\mu >0$. We have not found any qualitatively new features 
in points with negative $\mu$, as far as the determination of 
$m_{SS}$ is concerned.

Observables and their theoretically forecasted errors are taken from
the tables (5.13) and (5.14) of \cite{Weiglein:2004hn} and from
\cite{AguilarSaavedra:2005pw}. For the LHC we take into account the
``edge variables'': $(m_{ll})^{edge}$,
$(m_{lq})^{\text{edge}}_{\text{low}}$,
$(m_{lq})^{\text{edge}}_{\text{high}}$, $(m_{llq})_{\rm edge}$ and
$(m_{llq})_{\rm thresh}$ from the decay chain ${\tilde q}_L\to
\chi^0_2 q$ and $\chi^0_2 \to l {\tilde l} \to ll \chi^0_1$
\cite{Bachacou:1999zb,Allanach:2000kt,Lester:2001zx}. In addition, we
consider $(m_{llb})_{thresh}$, $(m_{\tau^+\tau^-})$ (from decays
involving the lighter stau) and the mass differences $\Delta_{{\tilde
g}{\tilde b}_i}= m_{\tilde g} - m_{{\tilde b}_i}$, with
$i = 1,2$, $\Delta_{{\tilde q}_R\chi^0_1}= m_{{\tilde q}_R} -
m_{\chi^0_1}$ and $\Delta_{{\tilde l}_L\chi^0_1}=
m_{{\tilde l}_L} - m_{\chi^0_1}$. Since $m_{{\tilde
e}_R}\simeq m_{{\tilde \mu}_R}$ and $m_{{\tilde u}_R}\simeq m_{{\tilde
d}_R}\simeq m_{{\tilde c}_R}\simeq m_{{\tilde s}_R}$ applies for a
large range of the parameter space LHC measurements will not be able
to distinguish between the first two generation sfermions. 
\footnote{See however \cite{Allanach:2008ib,Buras:2009sg,Abada:2010kj}.}
This allows
us to define the masses $m_{{\tilde l}_L} = (m_{{\tilde e}_L} +
m_{{\tilde \mu}_L})/2$ and $m_{{\tilde q}_R} = (m_{{\tilde u}_R} +
m_{{\tilde d}_R} + m_{{\tilde c}_R} + m_{{\tilde s}_R})/4$ which will
be used from now on for the mass differences $\Delta_{{\tilde
q}_R\chi^0_1}$ and $\Delta_{{\tilde l}_L\chi^0_1}$.  As discussed
below in section \ref{subsec:errors}, especially 
$\Delta_{{\tilde g}{\tilde b}_i}$ and the edge variables are
important for the LHC analysis. For the ILC we assume that at least
$m_{\chi^0_1}$, $m_{{\tilde e}_R}\simeq m_{{\tilde \mu}_R}$ and
$m_{{\tilde e}_L}\simeq m_{{\tilde \mu}_L}$ are kinematically
accessible. In addition, whenever within the reach of
the ILC, we also take into account ${\tilde\tau}_1$, 
$\chi^0_2$, $\chi^+_1$ and ${\tilde t}_1$, which are, however, less
important. We also assume that the lighter Higgs, $h^0$, has been
found and its mass measured with an accuracy which depends on whether
the analysis is for LHC only or for LHC+ILC, see the corresponding
error estimates in \cite{AguilarSaavedra:2005pw}. Errors for the ILC
are taken directly from the tables of the above papers.  For the error
bars for the LHC, however, we have rescaled all statistical errors
from the values for $300$ $fb^{-1}$ to a luminosity of (only) $100$
$fb^{-1}$. To be conservative the total error is
obtained summing statistic and systematic error linearly.  Note that
we did not make use of the combined LHC and ILC errors calculated in
the papers mentioned above. When we discuss the calculations for LHC
and ILC observables in \ref{subsec:LHCandILC} we refer to an analysis
in which the LHC and ILC observables are all enabled but the errors
are the errors for the LHC or ILC only. Nevertheless,  we have 
checked that using the combined errors changes the results  
only by an irrelevant amount. We will call this analysis therefore 
``ILC+LHC combined''.

In \cite{Weiglein:2004hn} and \cite{AguilarSaavedra:2005pw}, only 
standard SPS points have been studied in detail. In the calculation 
of the $\chi^2$-distributions we assume that {\em relative} errors 
for different mSugra points and/or seesaw points are constant. 
The assumption to use constant {\em relative} errors in all of our 
calculations is, of course, a crucial simplification which has to be 
checked very carefully. However, we have chosen to do so for the 
following two reasons: (a) It allows us to perform a $\chi^2$ 
analysis for all different spectra within reasonable CPU time. And 
(b)  uncertainties of the theoretically forecasted error bars 
are nearly impossible to estimate. 
Only experiments can finally determine total errors on observables. 
We thus use errors-as-predicted and discuss in section \ref{subsec:errors}, 
how our conclusions will change as a function of these unknown 
errors.

To numerically estimate the allowed ranges for the model parameters 
we use a simple $\chi^2$ procedure. We have found that, see below, 
errors on $m_0$, $M_{1/2}$ and $m_{SS}$ are very strongly correlated. 
To assure that our estimates are reliable in all cases we have written 
two completely independent numerical codes. 
The first of these is based on MINUIT, \footnote{Minimization package 
from the CERN Program  Library. Documentation can be found at
  \texttt{http://cernlib.web.cern.ch/cernlib/}}.
  enforcing the $m_{SS}$ scan while MINUIT is fitting the
parameters $m_0$, $M_{1/2}$, tan$\beta$ and $A_0$ for fixed $m_{SS}$.
The second code uses a straight-forward 
but slow Monte Carlo random walk procedure, which can be ``heated'' 
to find separated minima. In the MC calculations we use usually a 
(few) $10^{6}$  points to assure convergence. This makes the 
MC code slow, but reliable. We have done calculations using both 
codes in all cases necessary, to ensure that convergence has been 
reached. 

Finally we need to mention that in all calculations shown below we 
put neutrino Yukawa couplings to negligibly small values, unless 
noted otherwise. Again the reason for this choice is simply to 
limit the amount of CPU time necessary for our fits. \footnote{We 
let 5 parameters flow freely. For a type-II, for example, we have 
in $Y_T$ six more complex parameters. A full fit would require 
minimizing $\chi^2$ for 5+12-3=14 parameters, which can not done 
for all spectra we need to consider within realistic amounts of CPU 
time.} 
We will comment, however, in section (\ref{subsec:errors}) 
on the differences expected with fits, where the Yukawas are chosen 
to fit neutrino data correctly. In general, if $m_{SS}$ is below, 
say, $10^{14}$ GeV the differences of the full fit to our calculation 
with negligible Yukawas is found to be completely irrelevant. If 
any hints of seesaw with $m_{SS}$ in the range [$10^{14},10^{15}]$ 
were indeed found in SUSY spectra, however, we expect that 
a full analysis would find results which differ by some ($10-30$) \% 
(depending on the exact value of $m_{SS}$) from our preliminary 
numbers.

\subsection{\label{subsec:spec}Mass spectra and LHC and ILC observables}

In this subsection we briefly summarize the differences in 
the calculated mass spectra of the different models and how 
this affects the LHC and ILC observables, which we use 
in our fits. In the following all numerical results shown 
in the figures have been calculated solving the full 2-loop 
RGEs numerically.  Moreover, we have taken into account the
1-loop thresholds of the seesaw particles at the seesaw scale
as described in \cite{Esteves:2010ff} and include the one-loop
contributions to the SUSY masses \cite{Pierce:1996zz}. We have also 
included the shifts of the gauge couplings to the $\overline{DR}$-scheme.
It is instructive to discuss some 
(semi-) analytical approximations at 1-loop order, which will 
allow to understand qualitatively the numerical results. We 
stress that none of the following approximations is used in 
any way in the numerical calculations.

The introduction of complete $SU(5)$ multiplets at a scale 
below the GUT scale changes the running of the gauge couplings. 
At 1-loop order the gauge couplings at the different scales 
are given as \cite{Hirsch:2008gh}
\begin{eqnarray}
\alpha_1(m_Z) &=& \frac{5 \alpha_{em}(m_Z)}{3 \cos^2\theta_W}, \hspace{1cm}
\alpha_2(m_Z) = \frac{\alpha_{em}(m_Z)}{\sin^2\theta_W}, \\  \nonumber 
\alpha_i(m_{SUSY}) &=& \frac{\alpha_i(m_Z)}{1- \frac{\alpha_i(m_Z)}
               {4 \pi} b_i^{SM} \log{\frac{m_{SUSY}^2}{m_Z^2}}}, \\ \nonumber
\alpha_i(m_{SS}) &=& \frac{\alpha_i(m_{SUSY})}{1- \frac{\alpha_i(m_{SUSY})}
               {4 \pi} b_i \log{\frac{m_{SS}^2}{m_{SUSY}^2}}}, \\ \nonumber
\alpha_i(M_G) &=& \frac{\alpha_i(m_{SS})}{1- \frac{\alpha_i(m_{SS})}
               {4 \pi} (b_i+\Delta b_i) \log{\frac{M_G^2}{m_{SS}^2}}}.
\end{eqnarray}
Here, $b^{SM}=(b_1,b_2,b_3)^{SM}=\textstyle (\frac{41}{10},-\frac{19}{6},-7)$  
for SM and $b=(b_1,b_2,b_3)^{MSSM}= \textstyle (\frac{33}{5},1,-3)$  
for MSSM. $m_{SS}$ denotes the seesaw scale, i.e. the mass of the 15-plet 
or the mass(es)\footnote{We assume that the 3 copies of 24-plets are 
degenerate. In principle, given enough accurately measured observables, 
it might be possible to drop this assumption.} of the 24-plets. For the 
case of the 15-plet one finds $\Delta b_i=7$ whereas for the case 
with three 24-plets one finds $\Delta b_i=15$, since each $24_M$ gives a 
$\Delta b_i=5$. It is easy to show that at the 1-loop level 
the GUT scale is not changed by the introduction of the complete 
$SU(5)$ multiplets. However, the $\Delta b_i\ne 0$ lead to a faster 
``running'' of the gauge couplings and thus to a larger value of 
$\alpha(M_G)$ compared to the mSugra case. For seesaw scales smaller 
than roughly $m_{SS} \sim 10^{9}$ GeV ($10^{13}$ GeV) one then encounters 
a Landau pol in $\alpha(M_G)$ for seesaw type-II (type-III) 
\cite{Esteves:2010ff}. This defines in each case a lower limit 
on the seesaw scale, if we insist on perturbativity.\footnote{Note, 
however, the SPheno never allows us to push the seesaw scales down 
to these limits. Convergence problems are usually encountered
 already for $\alpha(M_G) \gsim 0.25$ 
depending on the mSugra parameters.}

Gaugino masses evolve like gauge couplings:
\begin{eqnarray}
M_i(m_{SUSY}) = \frac{\alpha_i(m_{SUSY})}{\alpha(M_G)} M_{1/2}.
\label{eq:gaugino}
\end{eqnarray}
Eq. (\ref{eq:gaugino}) implies that the ratio $M_2/M_1$, which is 
measured at low-energies, has the usual mSugra value, but the 
relationship to $M_{1/2}$ is changed. I.e. since $\alpha(M_G)$ 
is larger in the seesaw case than in the standard pmSugra,  
$M_i$ are smaller in seesaw than in pmSugra.

For the soft mass parameters of the first two generations one obtains
\cite{Hirsch:2008gh}
\begin{eqnarray}\label{eq:scalar}
m_{\tilde f}^2  &=& m_0^2  +  
\sum_{i=1}^3  c^{\tilde f}_i \left(
\left(\frac{\alpha_i(m_{SS})}{\alpha(M_G)}\right)^2 f_i
 + f_i'
\right) M_{1/2}^2, \\ \nonumber
 f_i &=& \frac{1}{b_i} \left(
1 - {\left[1 + \frac{\alpha_i(m_{SS})}{4 \pi} b_i \log
\frac{m^2_{SS}}{m_Z^2}\right]^{-2}} \right), \\
   f_i' &=& \frac{1}{b_i+\Delta b_i} \left(
1 - {\left[1 + \frac{\alpha(M_G)}{4 \pi} (b_i + \Delta b_i) \log
\frac{M_G^2}{m_{SS}^2}\right]^{-2}} \right).
\end{eqnarray}
The various coefficients $c^{\tilde f}_i$ are given in table~\ref{tab:coeff}.
\begin{table}[ht]
\begin{tabular}{|c|ccccc|}
\hline
$\tilde f$ &   $\tilde E$ & $\tilde L$ &$\tilde D$ & $\tilde U$ & $\tilde Q$ \\
\hline
$c^{\tilde f}_1$ & $\frac{6}{5}$ & $\frac{3}{10}$ & $\frac{2}{15}$
                 & $\frac{8}{15}$ & $\frac{1}{30}$ \\
$c^{\tilde f}_2$ & 0 &  $\frac{3}{2}$ &  0 & 0 & $\frac{3}{2}$ \\
$c^{\tilde f}_3$ & 0 & 0 &  $\frac{8}{3}$ & $\frac{8}{3}$ & $\frac{8}{3}$ 
\\ \hline
\end{tabular}
\caption{Coefficients $c^{\tilde f}_i$  for 
eq.~(\ref{eq:scalar}).}
\label{tab:coeff}
\end{table}

In the limit $m_{SS}\to M_G$ the functions $f_i'$ go to zero and
one recovers the standard 
mSugra estimations for the sfermion masses. For any $m_{SS}$ below 
$M_G$, the contribution from $f_i$ are smaller than in the mSugra 
case, due to the prefactor which is always smaller than one. The contribution 
from the $f_i'$ can only partially compensate for this and thus, 
at low energies for a given pair of $m_0$ and $M_{1/2}$ one expects 
the sfermion masses to be smaller in seesaw than in pmSugra. It is 
important, however, to note that the different coefficients 
$c_i$ differ not only from sparticle to sparticle, but also are 
different for the same particle but different gauge groups. This 
observation is fundamentally the reason explaining the statement 
that accurate sfermion mass 
measurements will allow to distinguish pure mSugra from mSugra plus 
seesaw.\footnote{If all $c_i$ where the same, one could always 
fit the data by a simple rescaling of $m_0$ and $M_{1/2}$.} As 
in the case of pmSugra, coloured particles are expected to be 
heavier than non-coloured ones in seesaw. 

Before discussing the numerical results, we briefly 
comment on seesaw type-I. In type-I one only adds singlets to the 
MSSM particle content. Thus, $\Delta b_i=0$ $\forall i$ and there is 
no deformation of the spectrum with respect to mSugra due to the gauge 
part. The only change one expects for type-I is due to a different 
running of $m_L^2$, when Yukawas are taken into account. Since only 
$m_L^2$ is affected, most of the observables we have discussed above 
are not sensitive to the seesaw scale in type-I and our current 
analysis can not directly be applied to type-I seesaw.

\begin{figure}
\begin{center}
\begin{tabular}{cc}
\resizebox{80mm}{!}{\includegraphics{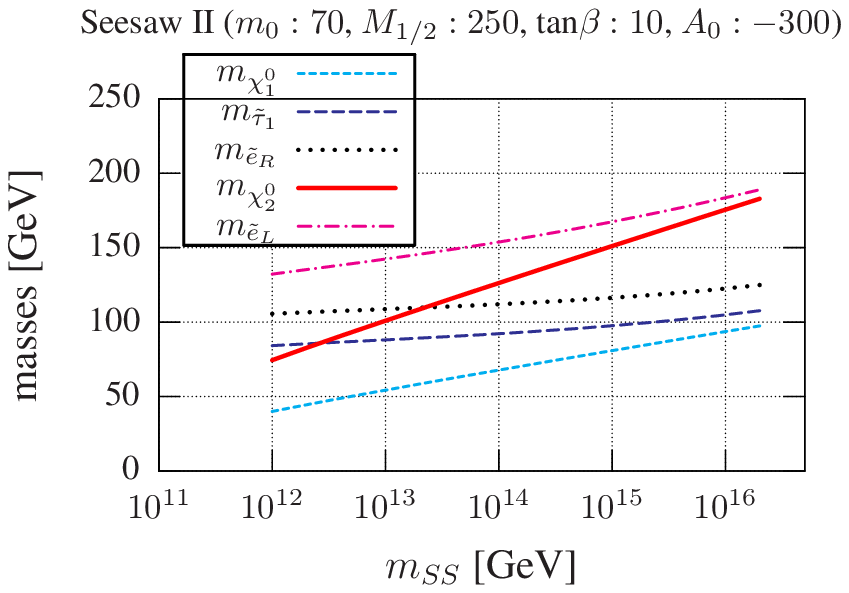}}    &
\resizebox{80mm}{!}{\includegraphics{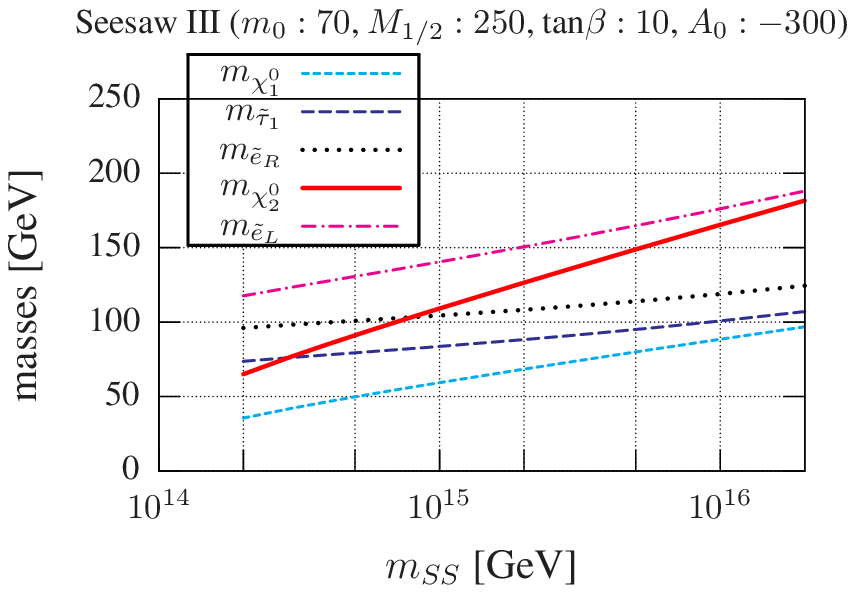}}   \vspace{2.5mm} \\
\resizebox{80mm}{!}{\includegraphics{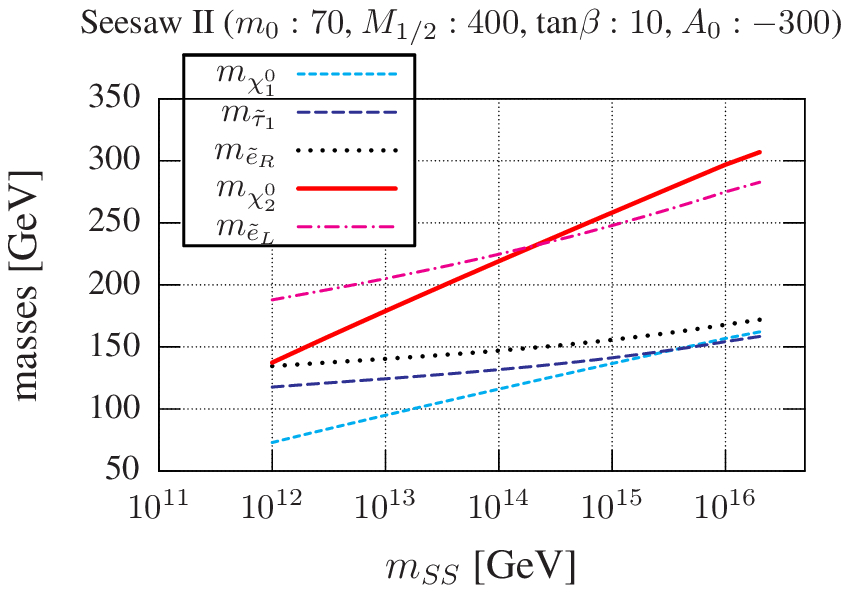}}   &
\resizebox{80mm}{!}{\includegraphics{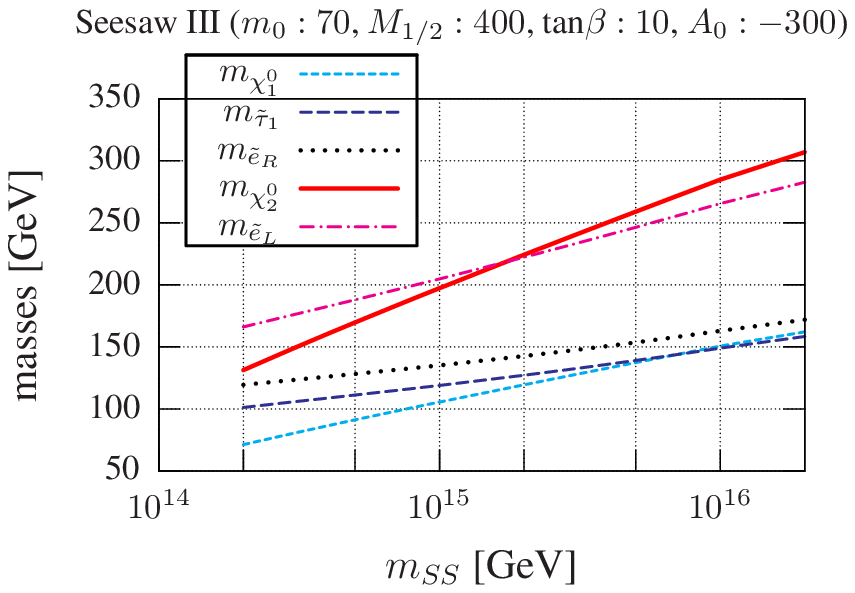}} 
\end{tabular}
\caption{\label{fig:running_masses_SPS1ap}Supersymmetric masses 
for two specific choices of mSugra parameters as a function of the 
seesaw scale. To the left: type-II. To the right: type-III. 
The mSugra parameters are fixed at the values of SPS1a' (top) 
and to MSP-1 (bottom). 
Note the different scales for type-II and type-III.} 
\end{center}
\end{figure}

Fig. (\ref{fig:running_masses_SPS1ap}) shows some examples of SUSY 
masses for two specific choices of mSugra parameters. The top panel 
shows mSugra parameters chosen as in the standard point SPS1a' 
\cite{AguilarSaavedra:2005pw}, while the bottom panel shows MSP-1 
for comparison. All plots show masses as function of $m_{SS}$, 
to the left for seesaw type-II and to the right for type-III. 
Shown are masses of the lighter two neutralinos, $\chi^0_1$ and 
$\chi^0_2$, and masses of charged sleptons. As is also typical in mSugra, 
the mass of the lighter chargino $m_{\chi^+_1}$ is very similar 
to $m_{\chi^0_2}$ and smuons and selectrons are nearly degenerate. 
We note in passing, that the mass of the lightest Higgs, $h^0$, 
shows little or no sensitivity at all on $m_{SS}$ but is important
in the fits. 

As discussed above, all masses get smaller for smaller values of 
$m_{SS}$ and are always smaller than in the pmSugra limit. Note the 
wide range for type-II shown and the much smaller range of $m_{SS}$ 
plotted for type-III. Ratios of gaugino masses follow standard 
expectations for all values of $m_{SS}$ and both types of seesaw. 
The slopes of the curves are different for different sparticles and 
the relative changes are larger in type-III than in type-II. This 
simply reflects the fact that type-III causes a larger change in 
the beta coefficients ($\Delta b_i=15$) than type-II ($\Delta b_i=7$). 

For the point SPS1a' the lighter chargino becomes lighter than 105 GeV 
for type-II (type-III) seesaw scales below roughly $m_{SS} \sim 
2\times 10^{13}$ GeV ($m_{SS} \sim 10^{15}$ GeV). Thus, $m_{SS}$ below these 
values are ruled out by the LEP bounds \cite{Nakamura:2010zzi,LEP2online}. 
Note that this implies that type-III can not explain neutrino data for 
mSugra parameters as in SPS1a' with Yukawas smaller than 
$Y_W^{ij} \le 1$.

\begin{figure}
 \begin{center}
  \begin{tabular}{cc}
   \resizebox{80mm}{!}{\includegraphics{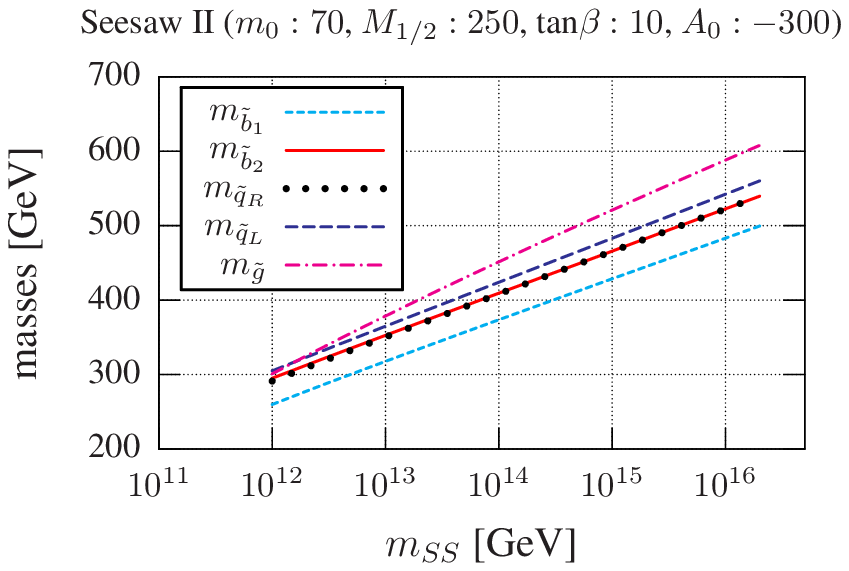}} &
   \resizebox{80mm}{!}{\includegraphics{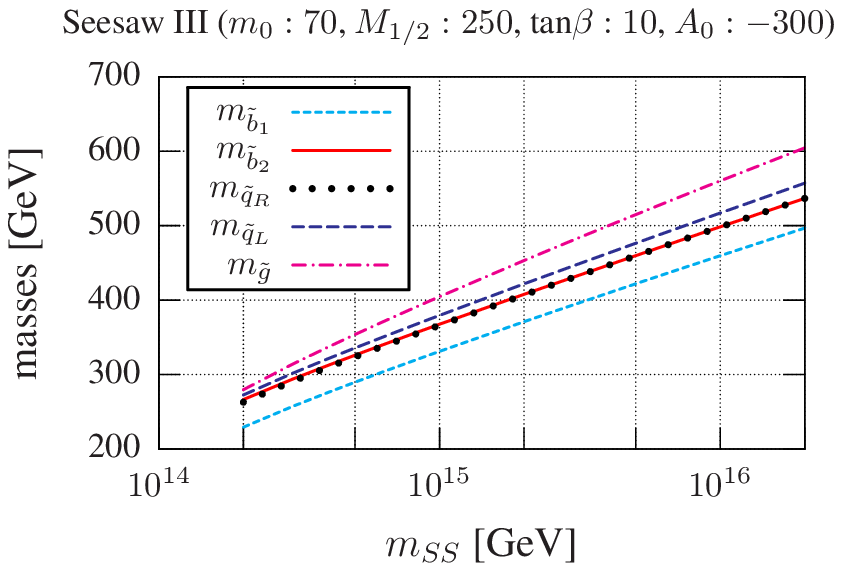}} 
  \end{tabular}
  \caption{\label{fig:running_masses_2} Running masses as a function of 
the seesaw scale, left: type-II; right: type-III. As fig. 
(\ref{fig:running_masses_SPS1ap}), but showing only the point SPS1a'. 
The masses shown in this figure are the most important coloured particles 
for our analysis.}
 \end{center}
\end{figure}

Changing the seesaw scale can lead to a different mass ordering for 
different sparticles. For example, for SPS1a' the $\chi^0_2$ is heavier 
than ${\tilde e}_R$ and ${\tilde\tau}_1$ in the mSugra limit (seesaw 
scale equal to $M_G$), but lighter than ${\tilde e}_R$ for type-II 
(type-III) seesaw scales below $m_{SS} \sim 3\times 10^{13}$ 
($m_{SS} \sim 8\times 10^{14}$). 
This is important for our study, since as a function of $m_{SS}$ it 
can happen that some observables are kinematically open for some 
values of $m_{SS}$ but not for others, see also the discussion in 
section \ref{subsec:LHC}.

The modified value of $M_{1/2}=400$ GeV in MSP-1 with respect to 
SPS1a' is motivated by the fact that for this choice of parameters 
the edge variables from the chain $\chi^0_2 \to {\tilde l}_R l\to 
l l \chi^0_1$ are kinematically possible for all relevant values of
$m_{SS}$. The larger value of $M_{1/2}$ implies heavier neutralinos 
and also that the LEP bounds on sparticle masses are fulfilled for 
all values of $m_{SS}$ shown. Note, that all sparticles shown in the 
plot are kinematically accessible at an ILC with $\sqrt{s}=1$ TeV.
For MSP-1 the lighter stau is the LSP for $m_{SS}$ larger than 
roughly $m_{SS}=10^{16}$ GeV. Thus, this point formally has no 
cosmologically acceptable pmSugra limit.

Fig.\ (\ref{fig:running_masses_2}) shows the dependence of several 
coloured sparticle masses on $m_{SS}$. Again to the left (right) we 
show seesaw type-II (type-III). Note the different scales 
for type-II and type-III. We show only the values for SPS1a' in 
this figure, masses for MSP-1 are larger but behave qualitatively 
very similar. The relative change of masses  as a function of 
$m_{SS}$ is much larger than for the non-coloured sparticle masses 
shown in fig.\ (\ref{fig:running_masses_SPS1ap}).  Here the range
where $m_{\tilde g} \simeq m_{\tilde q_{L,R}} \lsim 300$~GeV is excluded
by Tevatron data \cite{Nakamura:2010zzi,TevatronBounds}. However, this region 
is already excluded by LEP data.   Note 
that $m_{\tilde g} > m_{{\tilde b}_1}$ for all values of $m_{SS}$ 
in this point. Coloured sparticle production gives the bulk of 
the SUSY cross section at the LHC as usual. In these points most 
of the coloured sparticles are not kinematically accessible at 
the ILC, except for low values of $m_{SS}$. Except $\tilde{t}_1$ 
we therefore do not take into account measurements of coloured sparticles at 
the ILC in our analysis, even though they could be potentially 
much more accurate than the corresponding measurements at the 
LHC.

\begin{figure}
\begin{center}
\begin{tabular}{cc}
  \resizebox{80mm}{!}{\includegraphics
{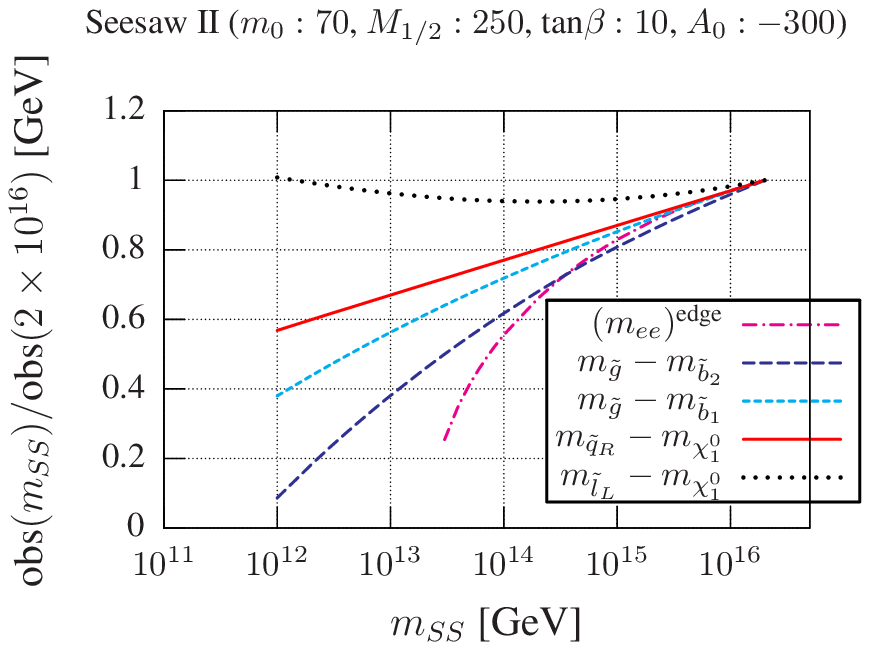}} &
  \resizebox{80mm}{!}{\includegraphics
{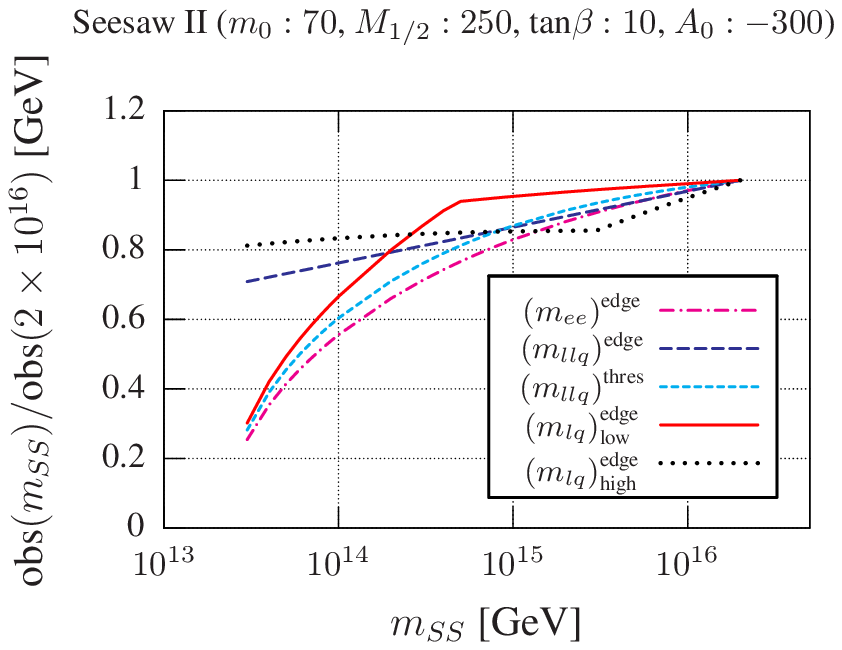}} \vspace{2.5mm} \\
  \resizebox{80mm}{!}{\includegraphics
{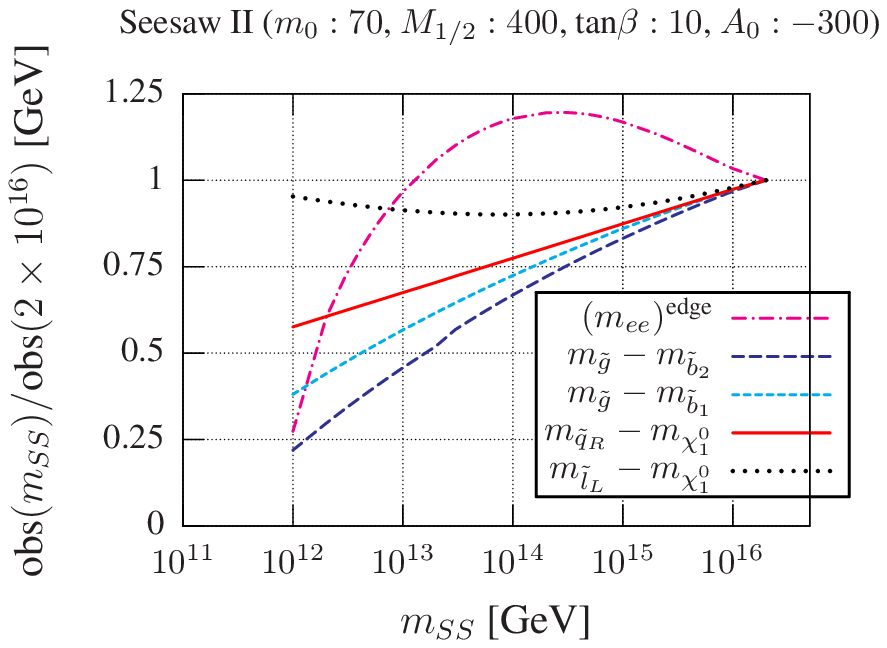}} &
  \resizebox{80mm}{!}{\includegraphics
{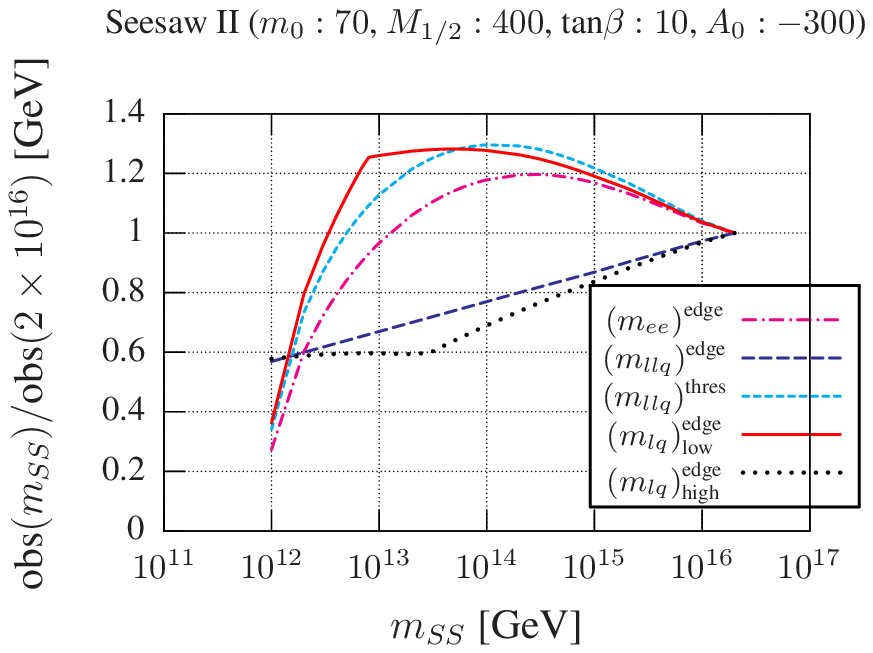}}
\end{tabular}
\caption{\label{fig:running_LHCobervables}Relative change of different LHC 
observables as a function of the seesaw scale for type-II seesaw. Top 
mSugra parameters as in SPS1a', bottom MSP-1. For an explanation see 
text. $(m_{ee})^{\rm edge}$ is repeated in the left plot for comparison.} 
\end{center}
\end{figure}

With the masses shown in fig. (\ref{fig:running_masses_SPS1ap}) and 
fig. (\ref{fig:running_masses_2}) one obtains the LHC observables 
shown in fig. (\ref{fig:running_LHCobervables}) for type-II. Again in the 
top panel we show SPS1a' and in the bottom panel MSP-1. The figure 
shows several mass differences (left) and the edge variables (right) 
stemming from the decay chain ${\tilde q} \to q \chi^0_2$ with the subsequent 
decay $\chi^0_2 \to l^{\pm} {\tilde l}^{\mp} \to l^{\pm}l^{\mp}\chi^0_1$ 
\cite{Bachacou:1999zb,Allanach:2000kt,Lester:2001zx}. 
We have normalized all observables to their expected values for 
$m_{SS}=M_G$. Thus relative changes in the different observables with 
respect to pmSugra are plotted. 

The two kinks in the running of $(m_{lq})^{\text{edge}}_{\text{low}}$
and $(m_{lq})^{\text{edge}}_{\text{high}}$ stem from the fact that one 
has to consider different cases in these observables. 
They  can be written as 
\cite{Allanach:2000kt}
\begin{eqnarray}\label{eq:edgecase}
(m_{lq})^{\text{edge}}_{\text{high}} &=& \text{max}[
(m_{l_{\text{near}}q}^{\text{max}})^2,
(m_{l_{\text{far}}q}^{\text{max}})^2 ]  \\ \nonumber
(m_{lq})^{\text{edge}}_{\text{low}} &=& \text{min}[
(m_{l_{\text{near}}q}^{\text{max}})^2, (m_{\tilde{q}}^2 -
m_{\chi^0_2}^2)(m_{\tilde{l}_R}^2 - m_{\chi^0_1}^2)/(2
m_{\tilde{l}_R}^2 - m_{\chi^0_1}^2) ]
\end{eqnarray}
where
\begin{eqnarray}\label{eq:edgecase2}
(m_{l_{\text{near}}q}^{\text{max}})^2 &=& (m_{\tilde{q}}^2 -
m_{\chi^0_2}^2)(m_{\chi^0_2}^2 - m_{\tilde{l}_R}^2)/m_{\chi^0_2}^2  \\ \nonumber
(m_{l_{\text{far}}q}^{\text{max}})^2 &=& (m_{\tilde{q}}^2 -
m_{\chi^0_2}^2)(m_{\tilde{l}_R}^2 -
m_{\chi^0_1}^2)/(m_{\tilde{l}_R}^2).
\end{eqnarray}
These conditions change as a function of $m_{SS}$ causing the kinks 
shown in the figure. Except for the $(m_{ll})^{\rm edge}$ different cases
appear in the expressions for all edges, but only for the variables 
$(m_{lq})^{\text{edge}}_{\text{low}}$ and
$(m_{lq})^{\text{edge}}_{\text{high}}$ do the kinematical conditions 
change as function of $m_{SS}$ normally.

The plot in fig. (\ref{fig:running_LHCobervables}) demonstrates the 
strong dependence of the LHC observables on $m_{SS}$. Increasing 
and decreasing values of the edges are possible, while mass differences 
usually decrease for lower values of $m_{SS}$. Note that in the 
$\chi^2$ fits, discussed in the next subsections, observables 
which show (a) the largest relative change with respect to $m_{SS}$ 
and (b) have the smallest expected errors will give the most important 
contributions. Finally we mention that fig. (\ref{fig:running_LHCobervables}) 
shows only type-II, since results for type-III are qualitatively 
similar but with larger relative changes.

\subsection{\label{subsec:LHCandILC}$\chi^2$ analysis for combined LHC and ILC 
data}

In this section we take into account all possible LHC and ILC 
observables. We discuss this more futuristic (but simpler) case 
first. Results for an analysis 
taking only LHC observables are discussed in the next subsection. 
We note in passing that we have checked that we can roughly reproduce 
the error on parameters for the pure mSugra results for the point SPS1a' 
discussed in detail in \cite{AguilarSaavedra:2005pw}. 

\begin{figure}
 \begin{center}
  \begin{tabular}{cc}
   \resizebox{75mm}{!}{\includegraphics
{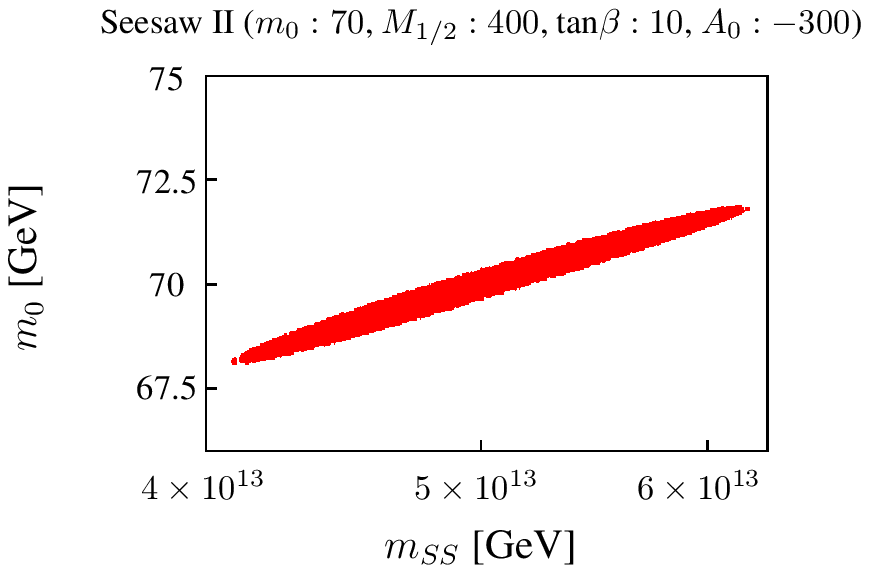}}   &
   \resizebox{75mm}{!}{\includegraphics
{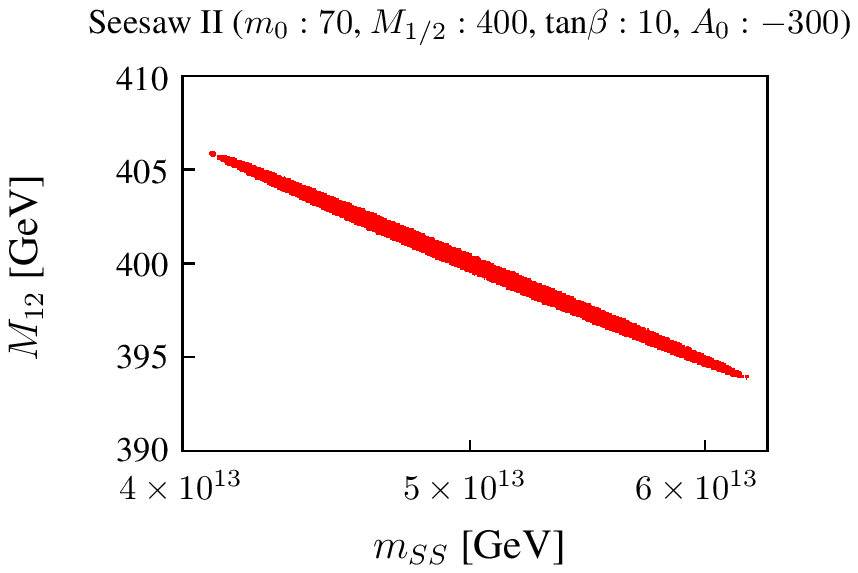}}  \\
   \resizebox{75mm}{!}{\includegraphics
{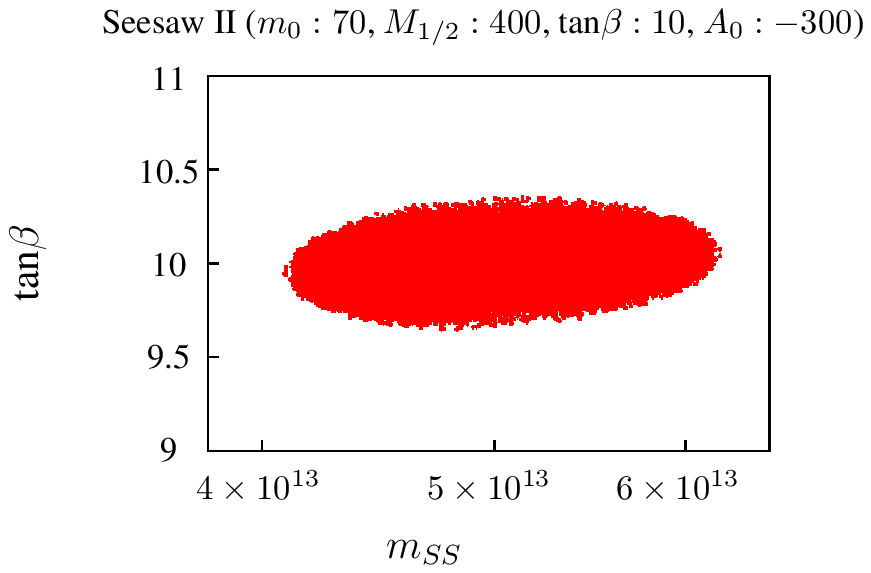}} &
   \resizebox{75mm}{!}{\includegraphics
{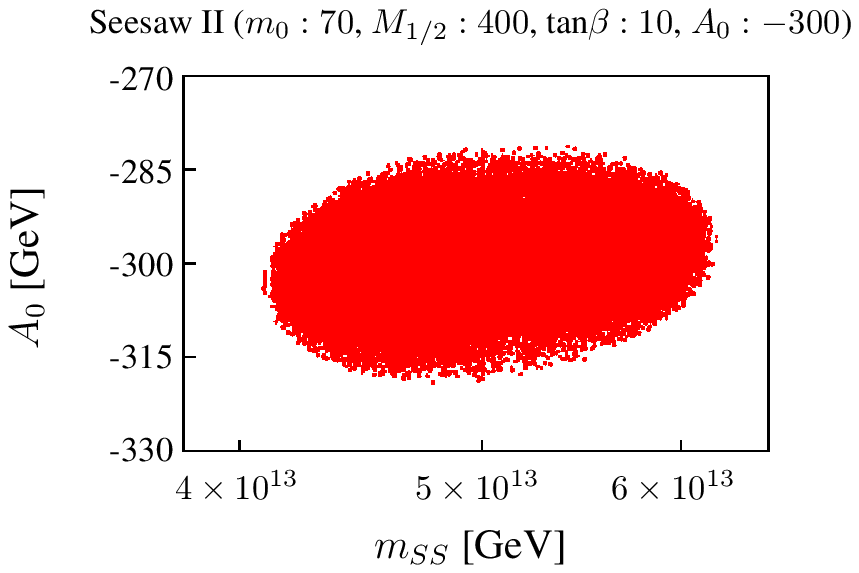}}   \\
   \resizebox{75mm}{!}{\includegraphics
{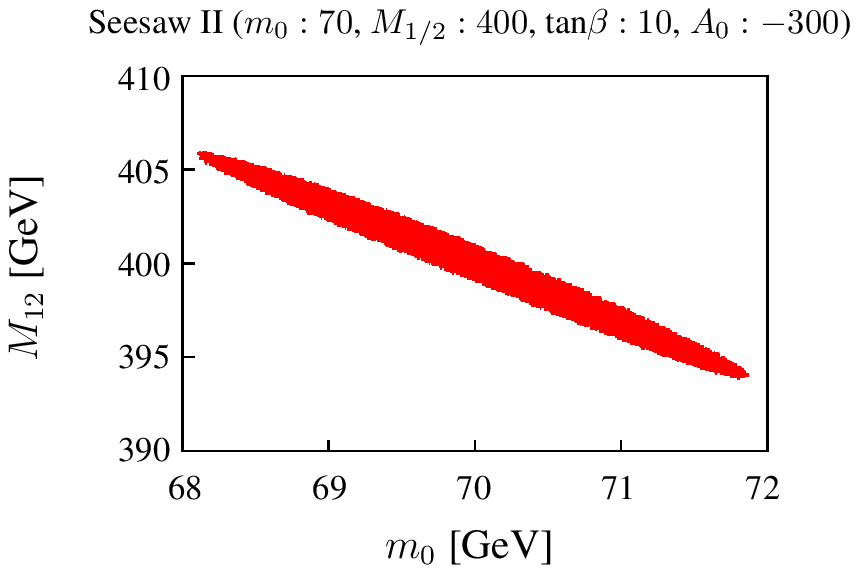}}     &
   \resizebox{76mm}{!}{\includegraphics
{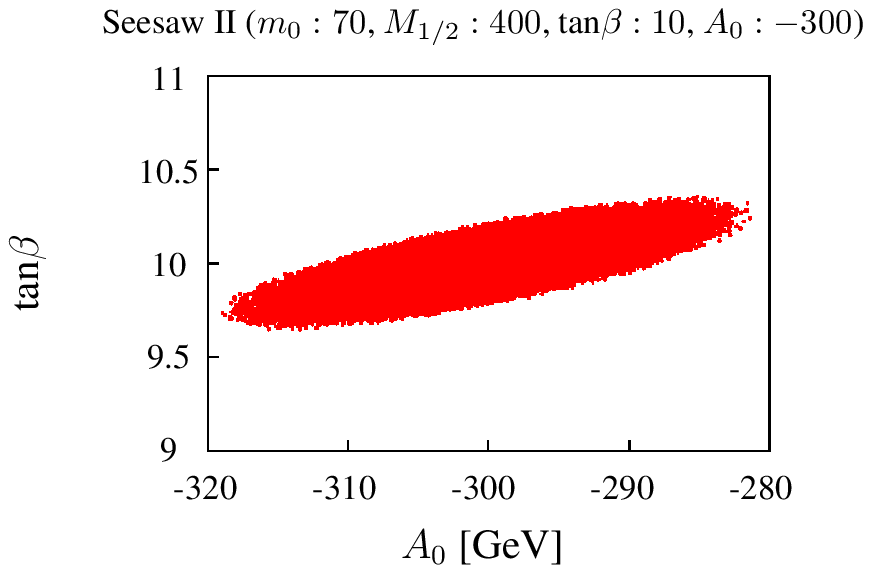}}    \\
  \end{tabular}
\caption{\label{fig:parameter_LHCandILC_SPS1App} Allowed parameter 
space for $m_0$, $M_{1/2}$, $\tan\beta$, $A_0$ and $m_{SS}$ for 
all 5 parameters varied freely. The input value for the seesaw scale is 
$m_{SS} = 5\times10^{13}$ GeV and seesaw type-II has been used.}
 \end{center}
\end{figure}

Fig. (\ref{fig:parameter_LHCandILC_SPS1App}) shows the allowed 
ranges of the parameters for the point MSP-1 and one specific choice 
of $m_{SS}=5\times 10^{13}$ GeV for type-II seesaw. The allowed regions 
have been found in a MC random walk procedure letting 5 parameters, 
$m_0,M_{1/2},A_0,\tan\beta$ and $m_{SS}$, float freely. The ranges shown 
correspond to a $\Delta\chi^2 \simeq 5.89$, i.e. 1 $\sigma$ c.l. for 
5 free parameters, where we have taken into account the 
correlations between the various parameters. 
Plotted are different 2-dimensional projections 
of parameters. 

As mentioned already above, the three parameters $m_0$, $M_{1/2}$ and 
$m_{SS}$ are highly correlated among each other. Lower values of 
$m_{SS}$ can be compensated by increasing $M_{1/2}$ and decreasing 
$m_0$ at the same time. This feature is present in all parameter space for 
both types of seesaw. This correlation results in errors on $m_0$ 
and $M_{1/2}$ which are larger (some times much larger, see below) 
than in pmSugra for the same input errors on observables. We note 
that the $\chi^2$ in this calculation is dominated by the much more 
accurate ILC data, see also the discussion in section (\ref{subsec:errors}) 
below.

In contrast, $\tan\beta$ and $A_0$ show very little correlation with 
$m_{SS}$ (and $m_0$ and $M_{1/2}$) and only a rather moderate correlation 
among themselves. $\tan\beta$ and $A_0$ are mostly determined by 
the Higgs mass measurement, and to  some extend by 3rd generation sfermions. 
Note that $A_0$ and $\tan\beta$ 
do not have much influence on the determination of $m_0$, $M_{1/2}$ and 
$m_{SS}$, apart from a slight increase in the errors of the latter. 
 However if $m_0$ cannot be fixed a determination
of $A_0$ and also  $\tan\beta$ becomes practically impossible, because
 almost any shift of $\tan\beta$ and $A_0$ can then be
compensated by changing  $m_0$ and/or $M_{1/2}$. 
This will be important when we
discuss the calculations using LHC observables only in section 
\ref{subsec:LHC}. 

For this choice of parameters, the error on $m_{SS}$ itself is found 
to be around $\Delta m_{SS} \sim 1.2 \times 10^{13}$, i.e. values of 
$m_{SS} = M_G$ are formally excluded by many standard deviations. 
However, $\Delta m_{SS}$ is a very strong function of $m_{SS}$ 
itself, as we will discuss below.

\begin{figure}
 \begin{center}
  \hspace{-7.5mm}
  \begin{tabular}{cc}
   \resizebox{76mm}{!}{\includegraphics{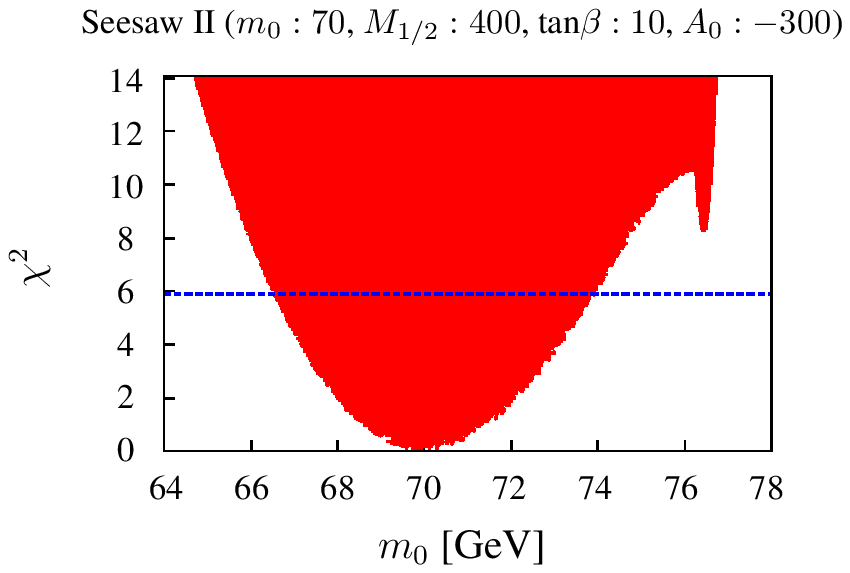}}   &
   \resizebox{75mm}{!}{\includegraphics{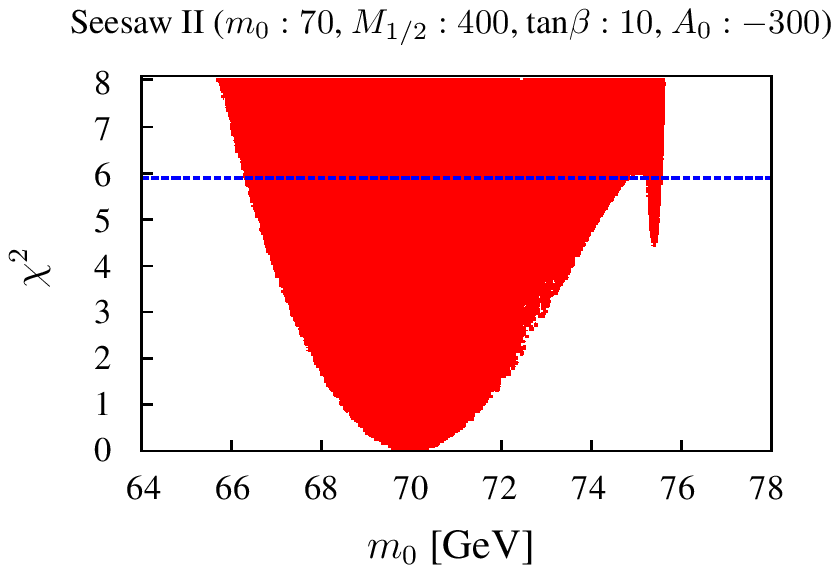}}   \\
   \resizebox{76mm}{!}{\includegraphics{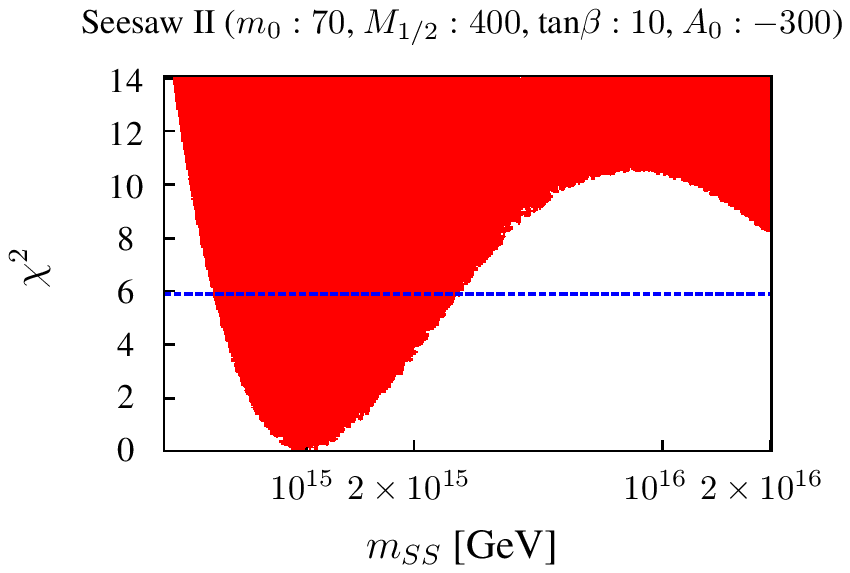}} &
   \resizebox{75mm}{!}{\includegraphics{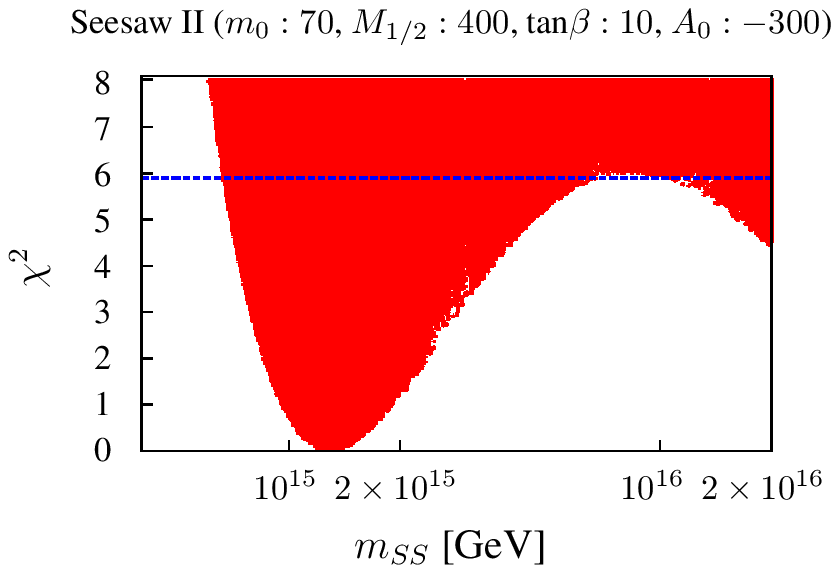}}
  \end{tabular}
  \caption{\label{fig:error_LHCandILC_SPS1App_ChiSq8and14_SII} $\chi^2$
  distributions of the random walk for MSP-1 and Seesaw type II. 
  The dashed line indicates a
  $\chi^2$ of $5.89$. Recall that $5.89$ corresponds to a $1\sigma$
  confidence level for five free parameters. The plots show the
  $\chi^2$ distributions for $m_{SS} = 1\times10^{15}$ GeV (left) and
  $m_{SS} = 1.3\times10^{15}$ GeV (right).}
 \end{center}
\end{figure}

Fig. (\ref{fig:parameter_LHCandILC_SPS1App}) shows results for a 
comparatively low value of $m_{SS}$. 
Fig. (\ref{fig:error_LHCandILC_SPS1App_ChiSq8and14_SII}) shows $\chi^2$
distributions for MSP-1 obtained by a random walk for seesaw type II 
and two slightly different but much higher values of $m_{SS}$: 
To the left: $m_{SS}=10^{15}$ GeV and to the right: $m_{SS}=1.3\times10^{15}$. 
The plots show the true $\chi^2$-minimum and a second (fake) minimum 
at $m_{SS}=M_G$. For the lower value of $m_{SS}$ this fake minimum 
is just excluded at 1 $\sigma$ c.l., while for the slightly higher 
value of $m_{SS}$ it is accepted at 1 $\sigma$ c.l. These kind of 
false minima appear in all our calculations when $m_{SS}$ approaches 
$M_G$. This is to be expected, since the models approach pmSugra 
in this limit. 

\begin{figure}
 \begin{center}
  \hspace{-7.5mm}
  \begin{tabular}{cc}
   \resizebox{76mm}{!}{\includegraphics{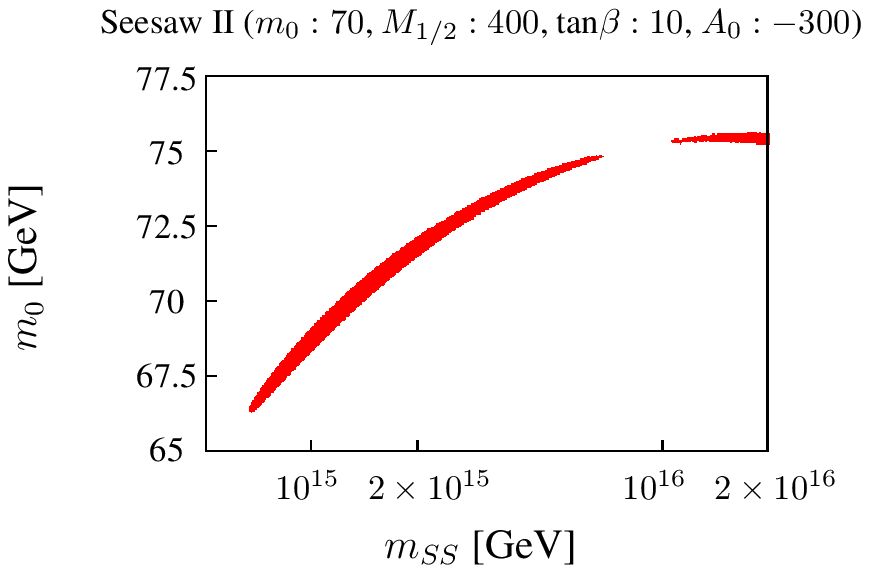}} &
   \resizebox{75mm}{!}{\includegraphics{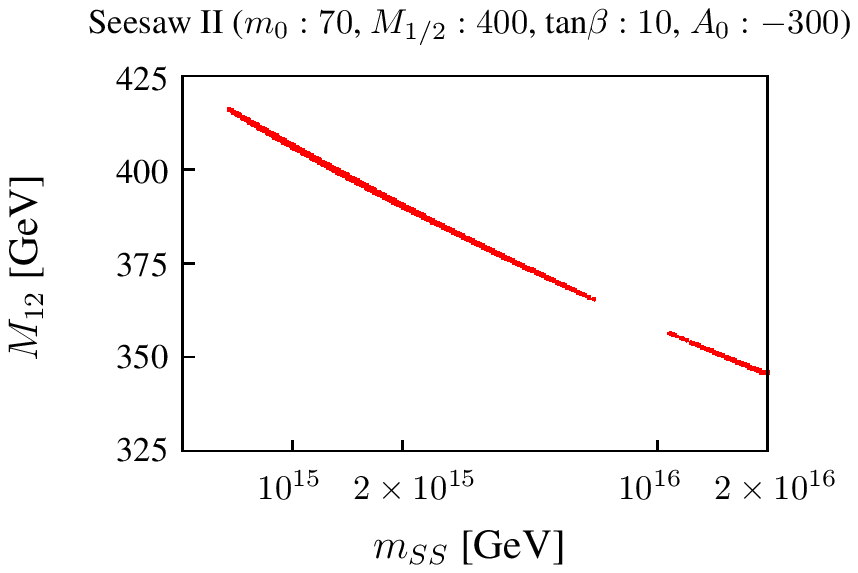}}
  \end{tabular}
  \caption{\label{fig:missing}Allowed ranges of parameters $m_{SS}$, 
  $m_0$ and $M_{1/2}$ for MSP-1 and type-II seesaw with input 
  $m_{SS}=1.3\times10^{15}$ GeV. Two separate solutions appear, one fake 
  but acceptable minimum is at $m_{SS}=M_G$.}
 \end{center}
\end{figure}

Fig.\ (\ref{fig:missing}) shows the allowed range of parameters 
$m_{SS}$, $m_0$ 
and $M_{1/2}$ for $m_{SS}=1.3 \times 10^{15}$ GeV. Two separate 
minima show up. For slightly larger values of $m_{SS}$ the two 
solutions overlap completely. Note that this also increases 
the errors on $m_0$ and $M_{1/2}$. For slightly smaller values of 
$m_{SS}$ this fake solution disappears resulting in a drastic 
decrease in the error bars of these three parameters. 
In case of type-III this fake minima does not show up separately, but 
indirectly by deforming the $\chi^2$ distributions. Thus for type-III 
the error bars go up to $M_G$ until it gets compatible with pmSUGRA. This 
will be important when we discuss mSUGRA plus type-III
later on.

In case of the ILC+LHC analysis this kind of ``false'' minima are usually 
the only class of fake minima that appear. Using only LHC data, the 
$\chi^2$ distributions are not that well behaved and false minima 
can also appear considerably below $M_G$, this will be shown in 
the next subsection.

\begin{figure}
 \begin{center}
  \hspace{-15mm}
  \begin{tabular}{cc}
   \resizebox{87mm}{!}{\includegraphics
{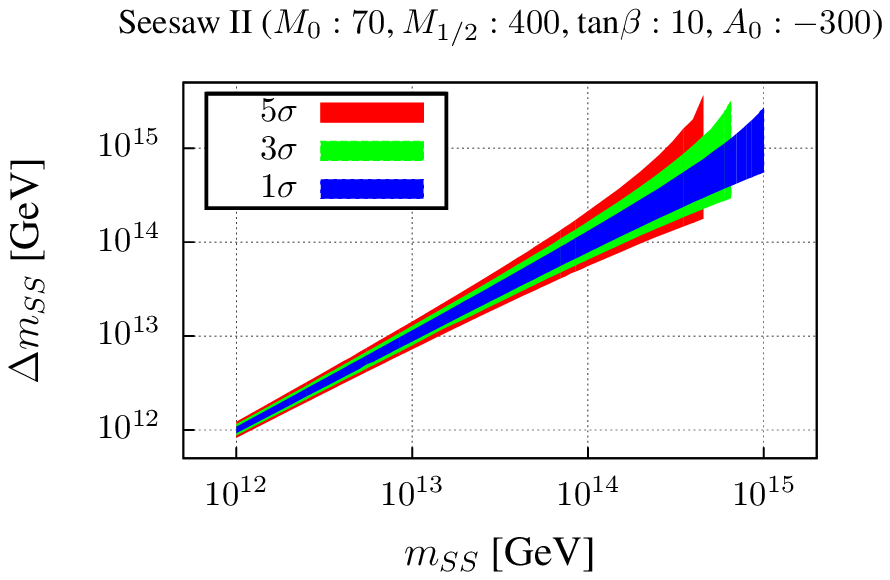}}  &
   \resizebox{80mm}{!}{\includegraphics
{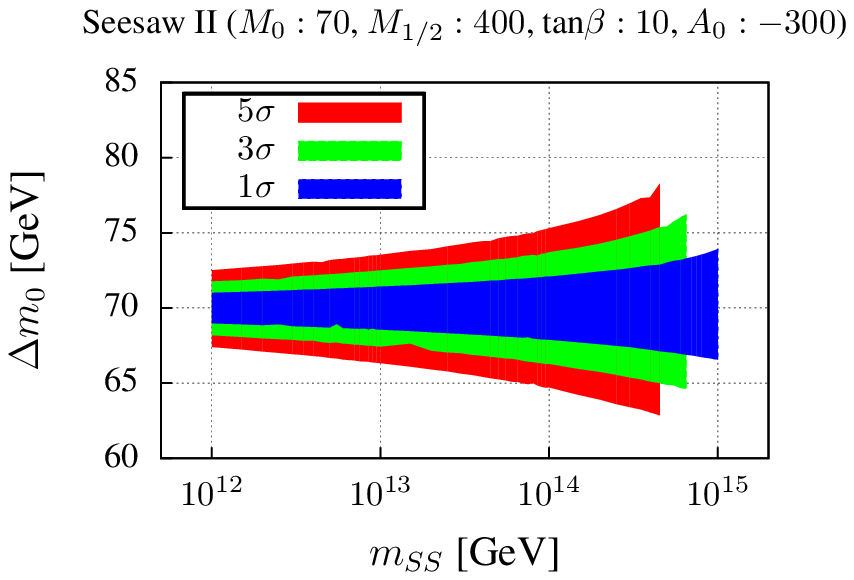}}     \\
   \resizebox{84mm}{!}{\includegraphics
{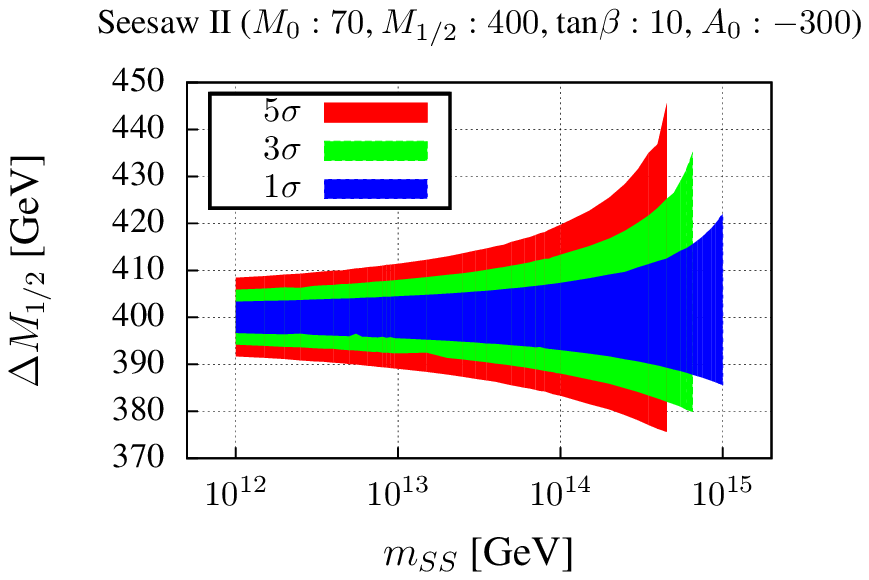}}    &
   \resizebox{83mm}{!}{\includegraphics
{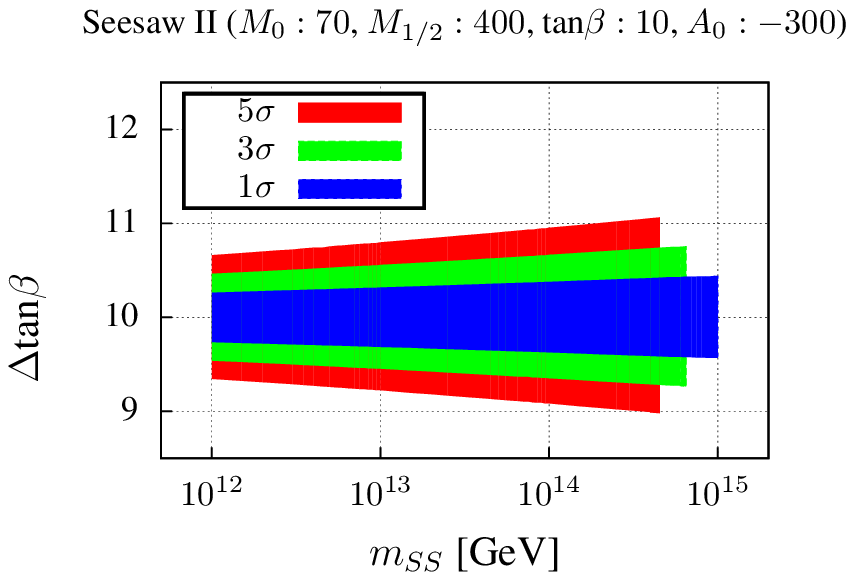}}   \\
   \resizebox{84mm}{!}{\includegraphics
{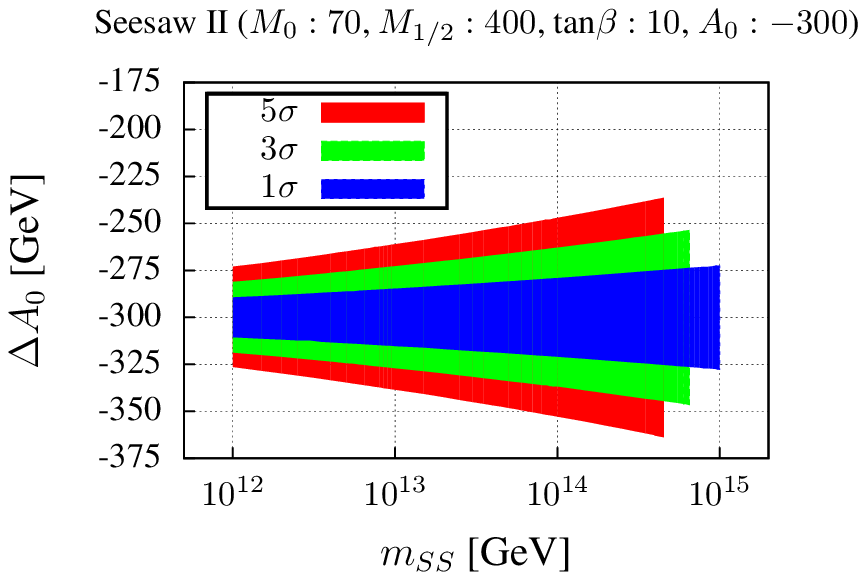}}    
  \end{tabular}
  \caption{\label{fig:error_LHCandILC_SPS1App_SSII} Error of $m_{SS}$,
  $m_0$, $M_{1/2}$, $tan\beta$ and $A_0$ against $m_{SS}$ for
  all 5 parameters freely varied. For these
  plots we used the LHC and ILC observables. The chosen values for
  $m_{SS}$, $m_0$, $M_{1/2}$, $tan\beta$ and $A_0$ are the values
  for MSP-1. The plots show the results for seesaw type
  II where we used a $1 \sigma$, $3 \sigma$ and $5 \sigma$ c.l..}
 \end{center}
\end{figure}

Fig. (\ref{fig:error_LHCandILC_SPS1App_SSII}) shows 1 $\sigma$, 
3 $\sigma$ and 5 $\sigma$ c.l. error bars on 
the different parameters of the model as a function of the seesaw 
scale for one specific mSugra set, MSP-1, for the case of type-II. 
The plots show a large range of $m_{SS}$ between 
[$10^{12},10^{15}$] GeV. Lower values of $m_{SS}$ are in principle possible, 
but show no new features. Larger values of $m_{SS}$ can not fit 
current neutrino data. Error bars on all parameters increase with 
increasing values of $m_{SS}$, and for values of $m_{SS}$ larger than 
(roughly) $(1-2)\times 10^{15}$ GeV the error $\Delta(m_{SS})$ is so large 
that the type-II can no longer be distinguished from pmSugra 
at the 1-$\sigma$ level, given the ILC+LHC observables with our 
``standard'' errors. $\Delta(m_{SS})$ decreases very rapidly as 
a function of $m_{SS}$ and for values of $m_{SS}= 6.5 \times 10^{14}$ 
($4.5 \times 10^{14}$) 
pmSugra and type-II can be formally distinguished by more than 3 (5) 
standard deviations.

Also the errors $\Delta(m_0)$ and $\Delta(M_{1/2})$ do show dependence on 
$m_{SS}$, especially at larger values of $m_{SS}$. Again, the reason 
for this dependence is found in the strong correlation among 
those three parameters, as discussed above. The error $\Delta(\tan\beta)$ 
(and to some extend $\Delta(A_0)$), on the other hand, shows less dependence 
on $m_{SS}$. This is explained by the fact that the lightest Higgs 
mass, $m_{h^0}$, shows very little dependence on the seesaw scale. 
 The slight dependence of $\Delta(A_0)$ on $m_{SS}$ is due to 
the lightest stop mass. For simplicity in all following plots we show 
only the 1 $\sigma$ allowed regions.

\begin{figure}
 \begin{center}
  \hspace{-7.5mm}
  \begin{tabular}{cc}
   \resizebox{80mm}{!}{\includegraphics
    {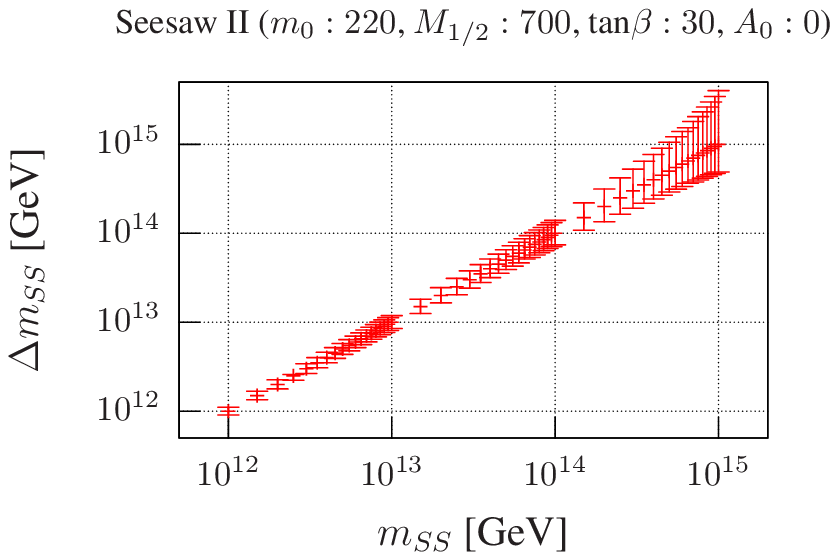}} &
   \resizebox{80mm}{!}{\includegraphics
    {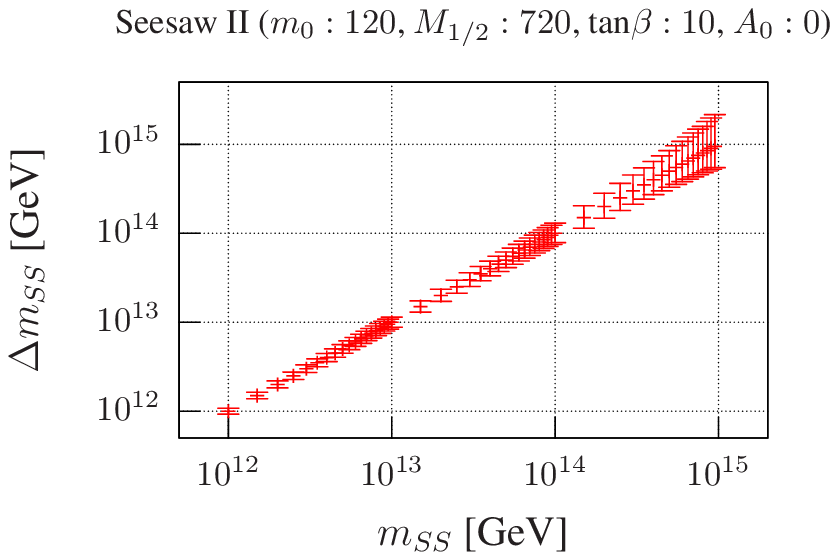}}
  \end{tabular}
\caption{\label{fig:error_LHCandILC_SSII} Error of $m_{SS}$,
  against $m_{SS}$ for 5 free parameters.  For these plots we used the
  LHC and ILC observables. The chosen values of parameters correspond 
  to MSP-2 (left) and MSP-3 (right).  The
  plots show the results for Seesaw type-II.}
 \end{center}
\end{figure}

Fig. ( \ref{fig:error_LHCandILC_SSII}) shows two more examples 
of $\Delta(m_{SS})$ as a function of $m_{SS}$. Here results for 
the points MSP-2 and MSP-3 are shown for type-II seesaw. Only 
$\Delta(m_{SS})$ as a function of $m_{SS}$ is shown. We do not 
repeat the plots for the other parameters because they are 
qualitatively very similar to the case shown in fig. 
(\ref{fig:error_LHCandILC_SPS1App_SSII}). As the plots show 
results for MSP-2 and MSP-3 are similar to MSP-1. Values of
$m_{SS}$ below roughly $m_{SS}\sim 10^{15}$ GeV are inconsistent 
with pmSugra. This implies that for the ILC errors as estimated 
in \cite{Weiglein:2004hn} and \cite{AguilarSaavedra:2005pw} a 
combined ILC+LHC analysis should be able to distinguish pmSugra 
from type-II seesaw for nearly all values of $m_{SS}$ relevant 
for neutrino data. We stress that this conclusion is correct only 
for those mSugra parameters for which (both left- and right-) 
sleptons and the lightest neutralino are kinematically 
accessible at the ILC.

\begin{figure}
 \begin{center}
  \hspace{-5mm}
  \begin{tabular}{cc}
   \resizebox{77mm}{!}{\includegraphics
    {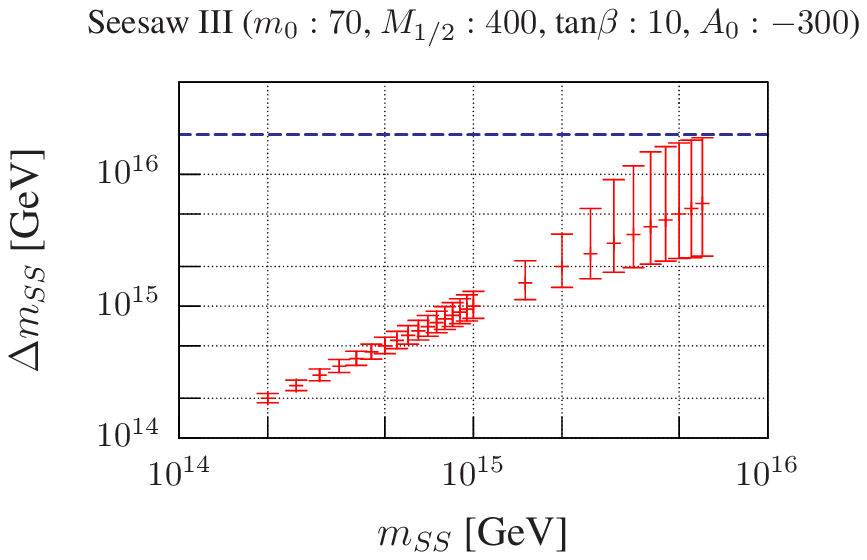}} &
   \resizebox{75mm}{!}{\includegraphics
    {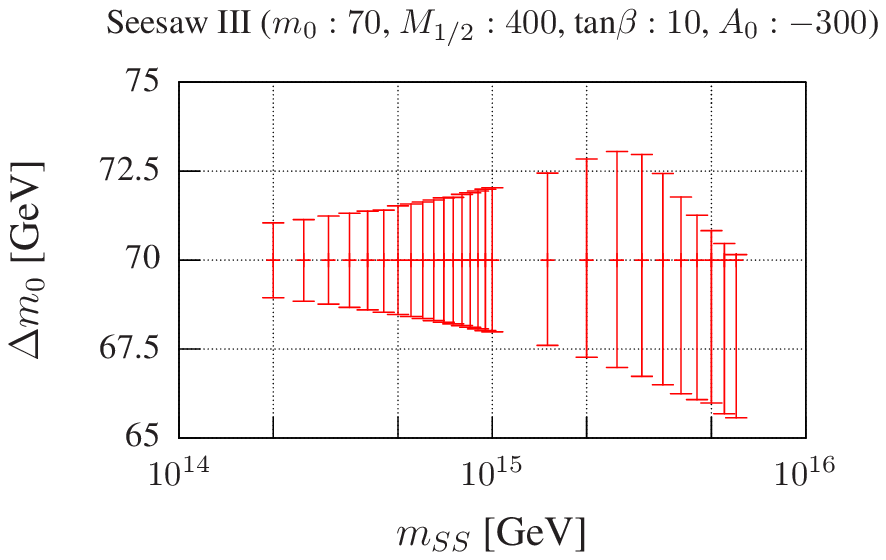}}    \\
   \resizebox{75mm}{!}{\includegraphics
    {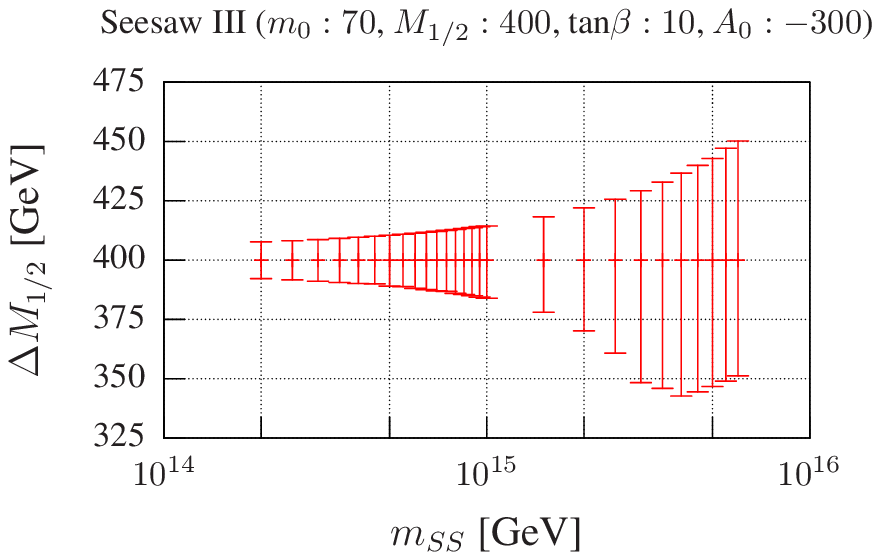}} 
  \end{tabular}
  \caption{\label{fig:error_LHCandILC_SPS1App_SSIII} Error of
  $m_{SS}$, $m_0$ and $M_{1/2}$ against $m_{SS}$ for 5 parameters 
  varied freely. For these plots we used the
  LHC and ILC observables. The chosen values for $m_{SS}$, $m_0$,
  $M_{1/2}$, $tan\beta$ and $A_0$ are the values according to
  MSP-1. The plots show the results for seesaw type III.}
 \end{center}
\end{figure}

Up to now we have shown only results for seesaw type-II. Fig. 
(\ref{fig:error_LHCandILC_SPS1App_SSIII}) shows a corresponding 
calculation for type-III and mSugra parameters as in MSP-1. 
Again, MSP-2 and MSP-3 show similar behaviour and we do not repeat 
the plots  for these points. Again, the scale of $m_{SS}$ is different 
from the case of type-II. Since SUSY masses show a stronger 
dependence on $m_{SS}$ in type-III than in type-II, larger values 
of $m_{SS}$ can be distinguished from pmSugra in this case. In the 
examples shown in the figure all values of $m_{SS}$ below roughly 
$m_{SS} \sim 5-6\times 10^{15}$ GeV can be distinguished from 
pmSugra with more than 1 $\sigma$ c.l. Recall that in type-III 
one expects $m_{SS} \lsim 8\times 10^{14}$ in order to explain 
neutrino data. Such ``low'' values of $m_{SS}$ differ from pmSugra 
in the fits by many standard deviations. 

Errors on $m_0$ are similar to the values observed for type-II, 
while $\Delta(M_{1/2})$ is larger in type-III. The correlation 
between $m_0$, $M_{1/2}$ and $m_{SS}$, discussed above for type-II, 
is also present in type-III and with an even larger correlation between 
$M_{1/2}$ and $m_{SS}$ in this case. The mSUGRA solution
does not show up explicitly as a second separate minimum, but 
 deforms the $\chi^2$ distributions, thus cutting the allowed 
ranges of $m_0$ and  $M_{1/2}$.

\begin{figure}
 \begin{center}
  \begin{tabular}{cc}
\resizebox{75mm}{!}{\includegraphics
{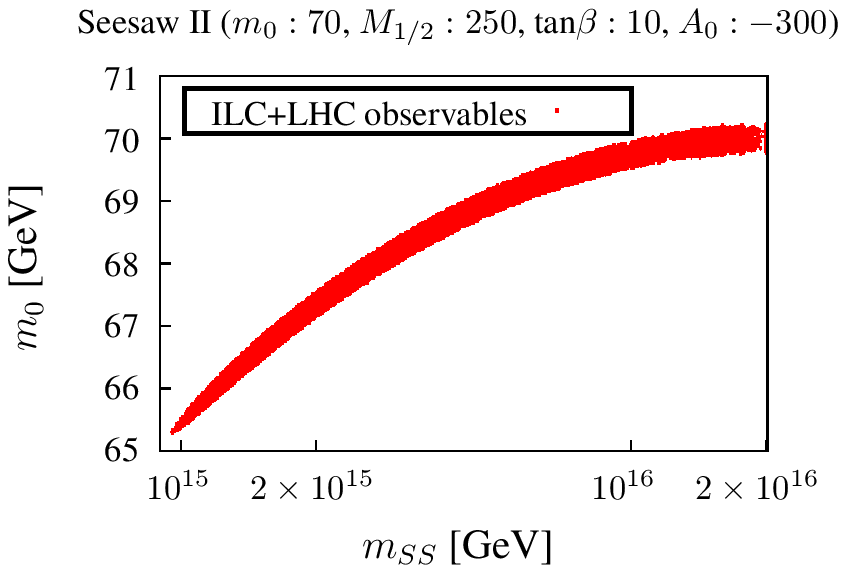}}   &
\resizebox{75mm}{!}{\includegraphics
{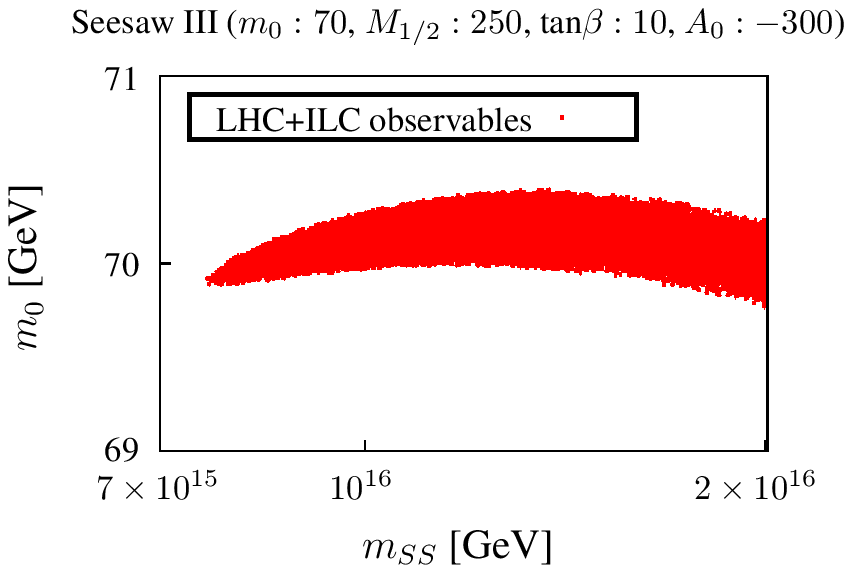}}
  \end{tabular}
  \caption{\label{fig:error_rwmSUGRA_SPS1Ap} The plots show random
  walks in which as starting point SPS1a' was chosen. For the
  parameter fit we used mSUGRA plus seesaw type-II and III, 
  respectively. The runs take into account LHC and 
  ILC observables. }
 \end{center}
\end{figure}

Up to now we have always used a seesaw spectrum as input. 
One can also ask the opposite question: Can a pmSugra 
point mimic a seesaw spectrum? An example of such a calculation 
is shown in fig. (\ref{fig:error_rwmSUGRA_SPS1Ap}). In this figure 
we show the allowed ranges for $m_{SS}$ and $m_0$ for mSugra parameters 
as in SPS1a' for type-II (left) and type-III (right). 
As one can see $m_{SS}$ as low 
as $m_{SS} \sim 10^{15}$ GeV ($m_{SS} \sim 7\times 10^{15}$ GeV) are allowed 
at 1 $\sigma$ c.l. for type-II (type-III) fits. Also note that 
$\Delta(m_0)$ is much larger than in a pmSugra fit, due again to 
the observed correlation among parameters. The results shown 
in fig. (\ref{fig:error_rwmSUGRA_SPS1Ap}) are consistent 
with the results discussed above, when a seesaw spectrum is used 
as input: $m_{SS}$ compatible with $M_G$ is reached at a very 
similar value of $m_{SS}$. 

Finally we note, that distinguishing type-II from type-III
requires extremely high precision, since they differ only 
at 2-loop order. The reason is that for 1-loop RGEs one can
always cancel the shifts in the coefficient of the beta-functions
by a rescaling of  $m_{SS}$. We have checked this numerically.

Closing this section we note that all results shown above have 
been obtained for the full 2-loop calculation. We have repeated 
the exercise in several cases using 1-loop RGEs only. As a general 
result, due to the weaker running of 1-loop RGEs, differences between 
pmSugra and seesaw are slightly smaller, leading to slightly larger 
errors on the parameters. For the case of the ILC+LHC analysis, however, 
differences between both calculations are rather small, with 
errors in parameters typically increasing in the order of (10-30) \% 
when going from a 2-loop to a 1-loop calculation.

\subsection{\label{subsec:LHC}LHC only}

In this subsection we discuss the results for an analysis using only 
LHC measurements. At the LHC observables do not measure SUSY masses 
directly. Instead, observables measure either mass differences or, 
in case of the edge variables, combinations of mass squared differences. 
Also one expects that LHC measurements will be much less precise than 
what can be done in case of the ILC. As a result the $\chi^2$ 
distributions for an LHC-only analysis show more complicated features 
than for the case of LHC+ILC. Especially it should be noted that in some 
cases we do not have a sufficiently large number of well determined 
observables and fake minima can appear, which will lead to sometimes 
rather large error bars on parameters, as discussed below.

\begin{figure}
 \begin{center}
  \hspace{-5mm}
  \begin{tabular}{ccc}
   \resizebox{76mm}{!}{\includegraphics
    {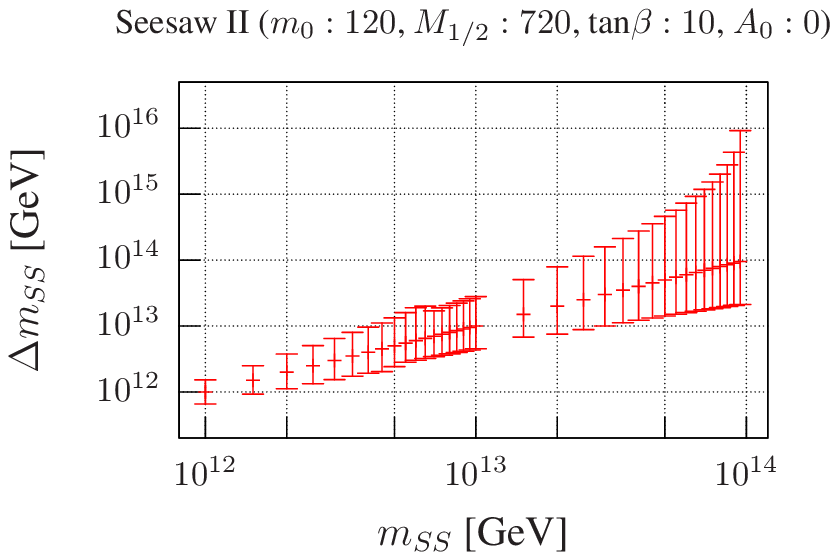}} &
   \resizebox{75mm}{!}{\includegraphics
    {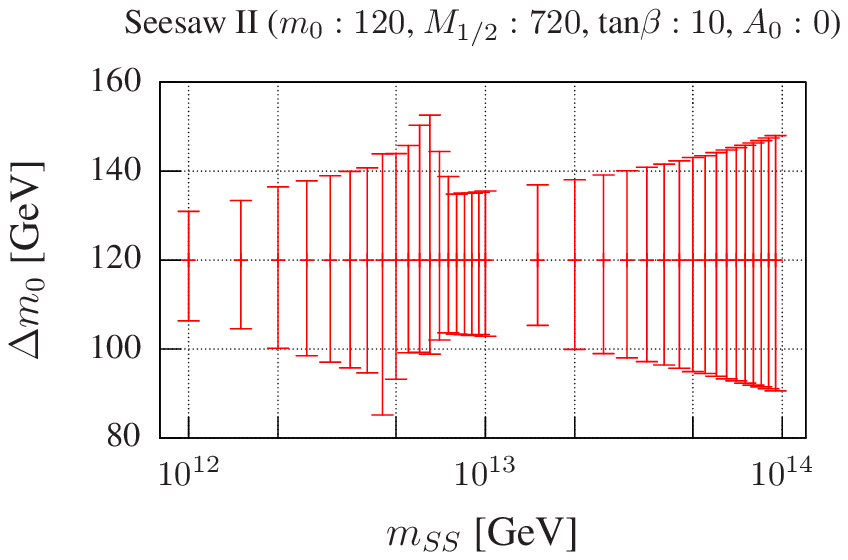}}    \\
   \resizebox{75mm}{!}{\includegraphics
    {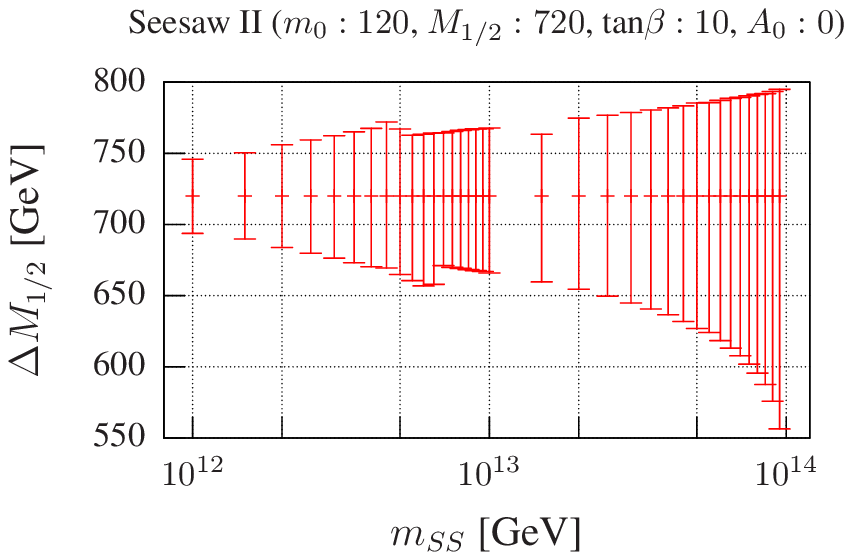}}
  \end{tabular}
  \caption{\label{fig:error_LHC_SPS3p_SSII} Error of $m_{SS}$, $m_0$,
  $M_{1/2}$, $tan\beta$ and $A_0$ against $m_{SS}$ for $m_{SS}$,
  $m_0$, $M_{1/2}$, $tan\beta$ and $A_0$ varied. For these plots we
  used only the LHC observables. The chosen values for $m_{SS}$,
  $m_0$, $M_{1/2}$, $tan\beta$ and $A_0$ are the values according to
  MSP-3. The plots show the results for seesaw type II.}
 \end{center}
\end{figure}

Fig. (\ref{fig:error_LHC_SPS3p_SSII}) shows error bars on $m_{SS}$, $m_0$, 
$M_{1/2}$ against $m_{SS}$ for the point 
MSP-3 and seesaw type-II, again for all 5 parameters varied freely. 
Note the change in the scale for $m_{SS}$, the largest value shown 
is $m_{SS}=10^{14}$ GeV. For larger values of $m_{SS}$ type-II 
seesaw can no longer be distinguished in this fit from pmSugra with 
at least 1 $\sigma$ c.l. Note, however, that $\Delta(m_{SS})$ decreases 
very rapidly for decreasing values of $m_{SS}$ and for values of 
$m_{SS}$ below $m_{SS}\sim$ (few) $10^{13}$ GeV pmSugra and type-II 
are formally different by several standard deviations.

The figure shows also that $\Delta(m_0)$ and $\Delta(M_{1/2})$ are 
much larger for the case of using only LHC observables than in the 
combined ILC+LHC analysis, as expected. Errors on $m_0$ and $M_{1/2}$ 
decrease in general with decreasing $m_{SS}$. The increase in $\Delta(m_0)$ 
and $\Delta(M_{1/2})$ around $m_{SS} \sim 7 \times 10^{12}$ GeV is due 
to the appearance of a fake side-minimum.  Such  fake minima
appear only for certain ranges of $m_{SS}$. Depending on which side
of the real minimum they appear they can lead to an asymmetric increase
of the errors as observed in this figure.

\begin{figure}
 \begin{center}
  \hspace{-5mm}
  \begin{tabular}{ccc}
   \resizebox{76mm}{!}{\includegraphics
    {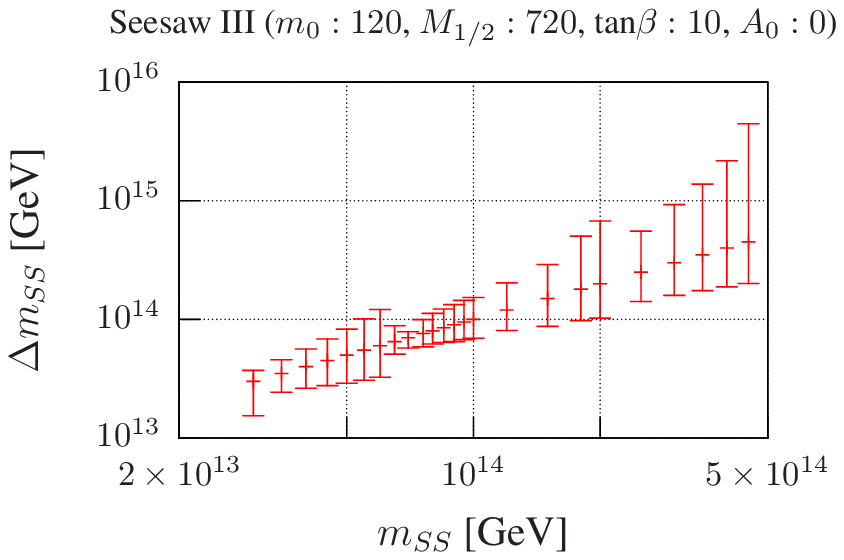}} &
   \resizebox{75mm}{!}{\includegraphics
    {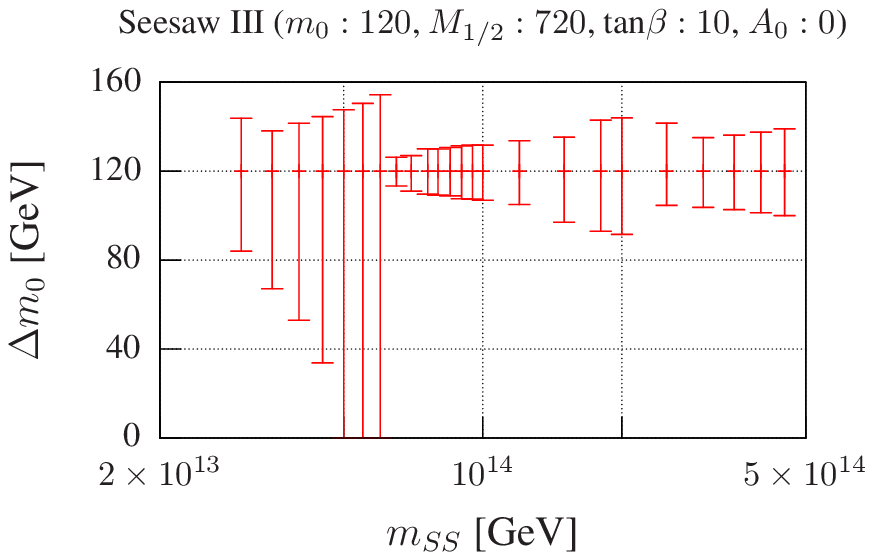}}   \\
   \resizebox{75mm}{!}{\includegraphics
    {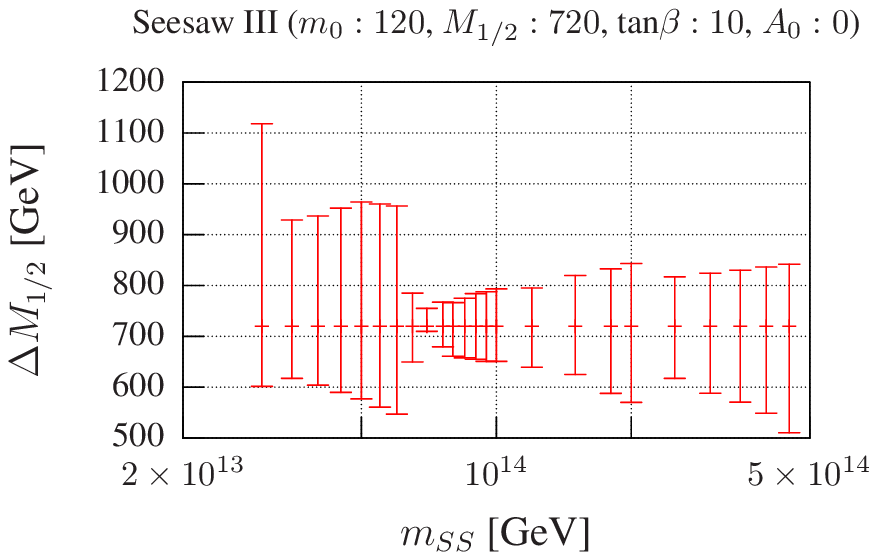}}
  \end{tabular}
  \caption{\label{fig:error_LHC_SPS3p_SSIII} Error of $m_{SS}$, $m_0$,
  $M_{1/2}$ against $m_{SS}$ for 5 parameters varied freely. For these plots 
  we used only the LHC observables. The chosen values for $m_{SS}$,
  $m_0$, $M_{1/2}$, $tan\beta$ and $A_0$ are the values according to
  MSP-3. The plots show the results for seesaw type III.}
 \end{center}
\end{figure}

Fig.\ (\ref{fig:error_LHC_SPS3p_SSIII}) shows an example of a corresponding 
fit for type-III. Again, $\Delta(m_0)$ and $\Delta(M_{1/2})$ and 
$\Delta(m_{SS})$ are shown as a function of $m_{SS}$ for MSP-3. Note that 
other points show qualitatively similar behaviour and that the range 
shown for $m_{SS}$ is comparatively small. Values of $m_{SS}$ larger 
than $m_{SS}\sim 5 \times 10^{14}$ are 1 $\sigma$ c.l. consistent with 
pmSugra. Since $m_{SS}\lsim 8 \times 10^{14}$ to explain neutrino 
data, LHC-only can probe interesting parts of the parameter space, 
but certainly will not be able to cover all possible values of 
$m_{SS}$ - unless LHC errors on mass measurements can be improved 
compared to expectations by considerable factors.

For decreasing $m_{SS}$ errors again decrease in general. There 
are two exceptions from this general rule in this plot. First, errors 
increase around $m_{SS} \sim 2\times 10^{14}$ GeV. This is again due 
to the appearance of a fake side minimum, 
which slowly disappears again when going towards smaller values of 
$m_{SS}$. The large increase in the error bars around $m_{SS} \sim 
6\times 10^{13}$ GeV is due to the fact that for smaller values of 
$m_{SS}$ in this calculation $\chi^0_2$ is lighter than ${\tilde e}_R$, 
i.e. the edges variables are lost completely. With only a few observables 
in the fit, all based on mass differences, $m_0$ and $M_{1/2}$ can 
hardly be fixed at all. The dramatic increase of the error
bars of $m_0$ can be understood easily from 
Eq.\ (\ref{eq:scalar}). As this equation shows the  sfermion masses behave 
approximately like $m_{\tilde{f}}^2 = m_0^2 + a M_{1/2}^2$. 
 When all edges are lost the remaining LHC 
observables can be fitted by varying $m_{SS}$ and  $M_{1/2}$ only.

\begin{figure}
 \begin{center}
  \begin{tabular}{cc}
\resizebox{75mm}{!}{\includegraphics
{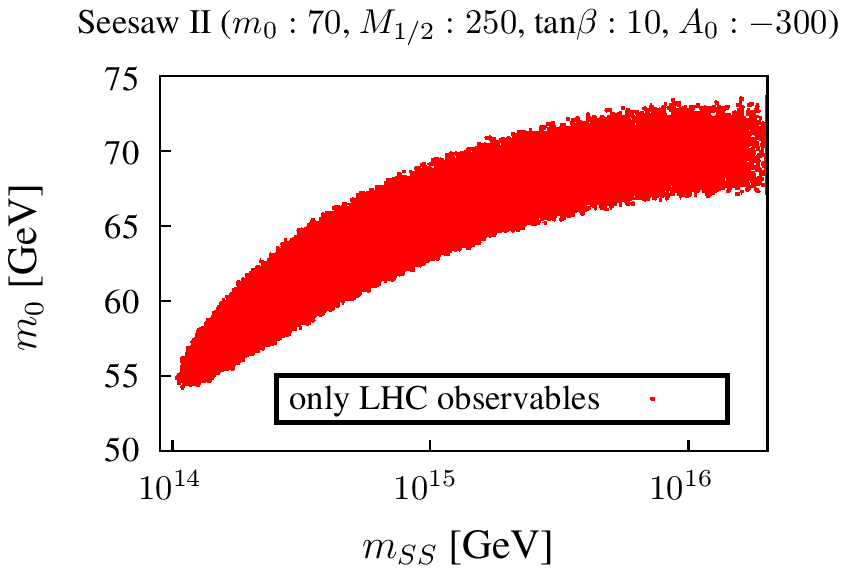}}   &
\resizebox{75mm}{!}{\includegraphics
{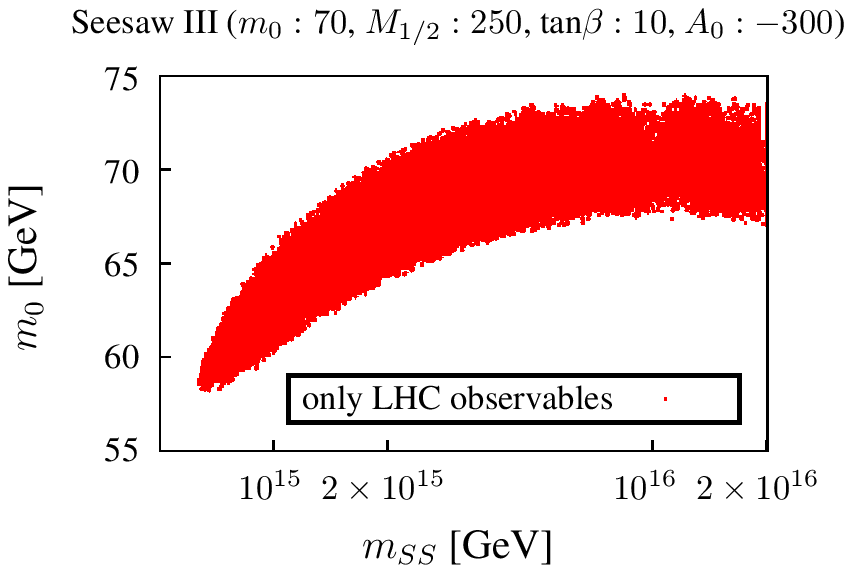}}  \\
  \end{tabular}
  \caption{\label{fig:error_rwmSUGRA_SPS1ApLHC} The plots show 
  the result of a random walk in which as a starting point SPS1a' 
  was chosen. For the parameter fit we used mSUGRA plus seesaw type II 
  and III, respectively. The runs were made for LHC observables enabled
  only.}
 \end{center}
\end{figure}

Finally we have calculated the allowed parameter space in a seesaw 
fit when the true input point is pmSugra. Two examples are shown 
in Fig. (\ref{fig:error_rwmSUGRA_SPS1ApLHC}). The mSugra parameters 
are for SPS1a' and type-II (type-III) seesaw is shown to the left 
(right). The allowed regions are much larger than in the combined 
ILC+LHC analysis, compared to the discussion in the last subsections. 
For type-II (type-III) values of $m_{SS}$ as low as 
$m_{SS} \lsim 10^{14}$ GeV ($m_{SS} \lsim 6 \times 10^{14}$ GeV) 
are allowed at the 1 $\sigma$ level. This is similar - and 
consistent - with the results discussed above for the opposite fit. 

In summary mass measurements from the LHC only should be able to 
distinguish between pmSugra and type-II (type-III) seesaw for seesaw 
scales below roughly $m_{SS} \lsim 10^{14}$ GeV ($m_{SS} \lsim  
6 \times 10^{14}$ GeV). This conclusion depends 
critically on the possibility to measure accurately several observables, 
as we are going to discuss next.

\subsection{\label{subsec:errors} Required accuracies on mass measurements 
and $\Delta(m_{SS})$}

All our results shown above crucially depend on the size of the 
expected error bars for the different observables. In this 
section we therefore discuss in some detail: (a) Which are the 
most important observables in our fits? And, (b) How accurately 
do we need to measure them to distinguish pmSugra from seesaw 
for a given, fixed value of $m_{SS}$. Again we will discuss the 
ILC+LHC case first.

\begin{figure}
 \begin{center}
  \begin{tabular}{cc}
\resizebox{75mm}{!}{\includegraphics
{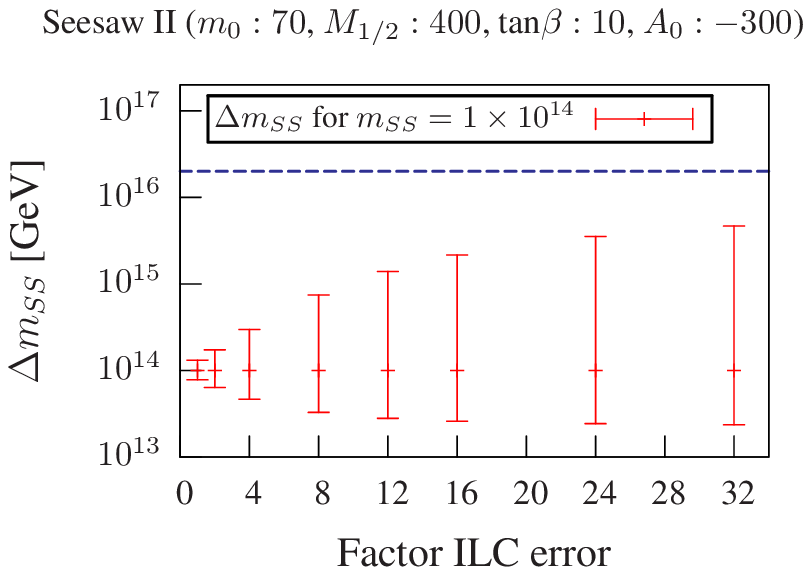}} &
\resizebox{75mm}{!}{\includegraphics
{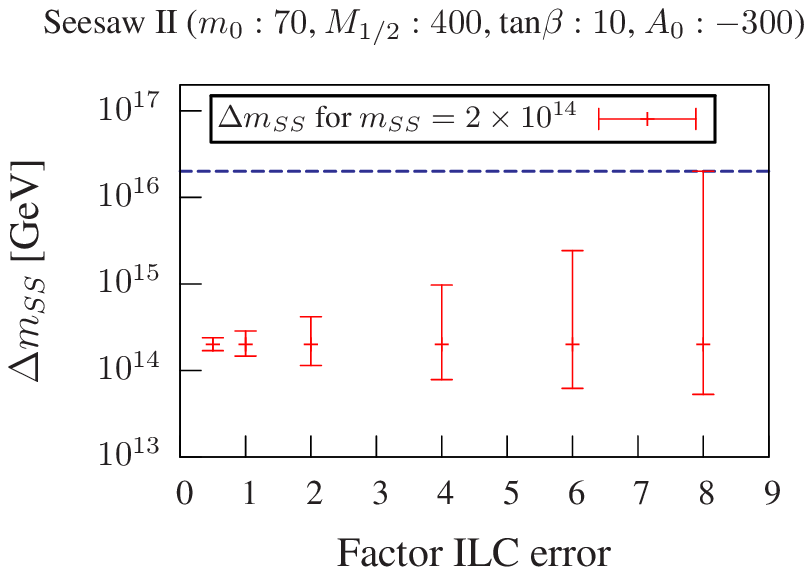}}
  \end{tabular}
  \caption{\label{fig:error_ILCobservables}Error of $m_{SS}$ with
  respect to the error of the ILC observables, $f_{ILC}$.}
 \end{center}
\end{figure}

Fig. (\ref{fig:error_ILCobservables}) shows $\Delta(m_{SS})$ for the 
points MSP-1 for two choices of $m_{SS}$. $\Delta(m_{SS})$ is shown 
as a function of the error of the ILC mass measurements. According 
to \cite{AguilarSaavedra:2005pw} it is expected that the ILC can 
measure SUSY masses of $\chi^0_i$ and ${\tilde l}$ states kinematically 
accessible with errors of the order (0.5-2) per-mille. We define a 
common factor $f_{ILC}$ and multiply all relative errors given in 
table 6 of \cite{AguilarSaavedra:2005pw} with this common factor. 
$\Delta(m_{SS})$ is then shown as a function of this factor in 
fig. (\ref{fig:error_ILCobservables}). Note that in this calculation 
we keep all LHC errors unchanged at their ``standard values''.

As fig. (\ref{fig:error_ILCobservables}) to the left shows, 
$\Delta(m_{SS})$ increases with the assumed errors of the ILC 
measurements. However, for MSP-1 and $m_{SS}=10^{14}$ GeV, LHC 
measurements alone are sufficient to distinguish type-II from 
pmSugra. Thus, error bars on $m_{SS}$ hardly increase going 
from $f_{ILC}=24$ to $f_{ILC}=32$. This means that ILC data dominate 
the fit until errors are about one order of magnitude larger 
than estimated in \cite{AguilarSaavedra:2005pw}, for larger ILC 
errors LHC measurements become more important for this choice 
of $m_{SS}$. 

The situation is quite different for $m_{SS}=2\times 10^{14}$ GeV, 
see fig. (\ref{fig:error_ILCobservables}) right. While  $f_{ILC}=6$
still allows to distinguish between pmSugra and type-II, for 
$f_{ILC}=8$, $\Delta(m_{SS})$ becomes to large 
to differentiate 
between type-II and pmSugra. The required accuracy of measurements 
of SUSY masses at the ILC is therefore a strong function of $m_{SS}$ 
itself. Errors of the order (1-2) percent are in general tolerable 
for seesaw scales below $m_{SS}=10^{14}$ GeV, while per-mille level 
measurements are required in the interval $[10^{14},10^{15}]$ GeV. 
We note that other SUSY points behave very similar and that for 
type-III correspondingly larger errors are tolerable.

\begin{figure}
 \begin{center}
  \resizebox{80mm}{!}{\includegraphics
   {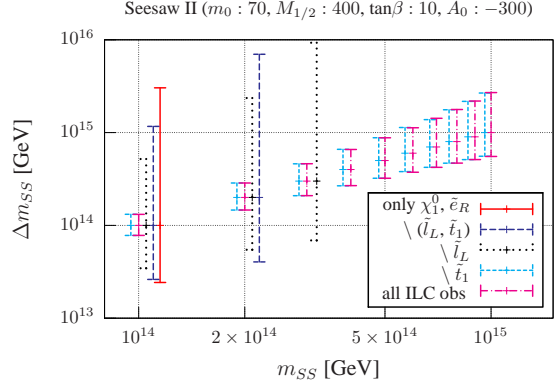}}
  \caption{\label{fig:error_LHCandILC_SPS1App_ObsONOFF_SII} In this
  plot the errors of $m_{SS}$ with respect to $m_{SS}$ are shown. The
  different lines belong to different runs where different combinations 
  of ILC observables were switched off. The different lines
  belong in each case to the same $m_{SS}$ values but were 
  a bit to be able to distinguish them. The real $m_{SS}$ value is the 
  value of the pink error bars which correspond
  to the run for all ILC observables enabled.}
 \end{center}
\end{figure}

Fig. (\ref{fig:error_ILCobservables}) treats all ILC observables equally. 
An interesting question to ask is, of course, which ILC observables 
are the most important ones for our analysis. Fig. 
(\ref{fig:error_LHCandILC_SPS1App_ObsONOFF_SII}) provides the answer.
Again for the point MSP-1 and for seesaw type-II we show $\Delta(m_{SS})$ 
as a function of $m_{SS}$ for different calculations taking into 
account different observables. We have kept all LHC observables 
``on'' at their standard errors. ``All ILC obs'' is the standard 
fit, taking into account all kinematically accessible mass measurements 
with their original errors from \cite{AguilarSaavedra:2005pw}. 
We then switched off by hand completely the contributions from different 
observables. Switching off the measurement of the mass of ${\tilde t}_1$ 
hardly changes the result. On the other hand, it can be seen that 
measuring left-slepton masses is highly important. Error bars increase 
sizeably if this observable is not taken into account and while 
a set of measurements with all observables can distinguish pmSugra 
from type-II all the way up to $m_{SS}=10^{15}$ GeV, without the 
accurate measurement of $m_{{\tilde l}_L}$ all values of 
$m_{SS}\gsim 4 \times 10^{14}$ GeV are compatible with pmSugra at the 1 
$\sigma$ level.  The relative importance of ${\tilde l}_L$ despite 
its larger error stems from the fact that $m_{{\tilde l}_R}$ has very 
little sensitivity to $m_{SS}$, see fig. (\ref{fig:running_masses_SPS1ap}).

\begin{figure}
 \begin{center}
  \begin{tabular}{cc}
\resizebox{75mm}{!}{\includegraphics
{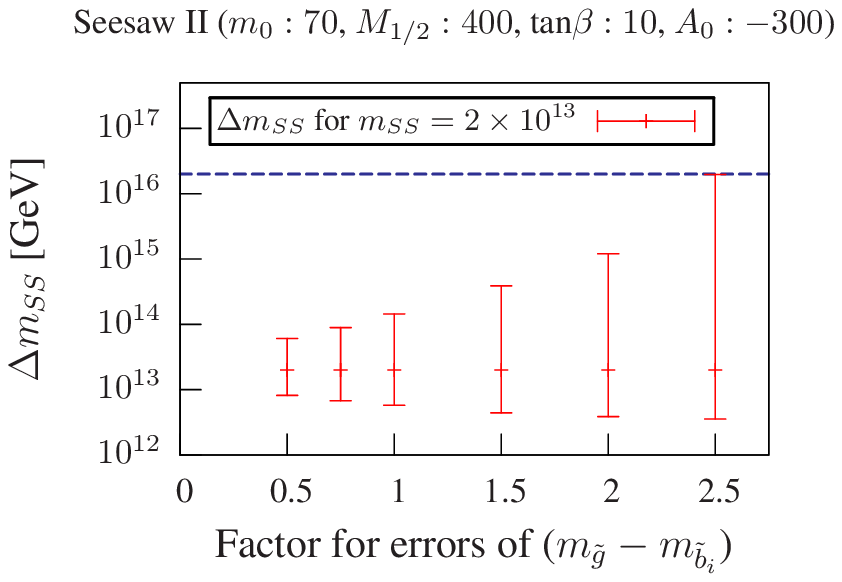}} &
\resizebox{76mm}{!}{\includegraphics
{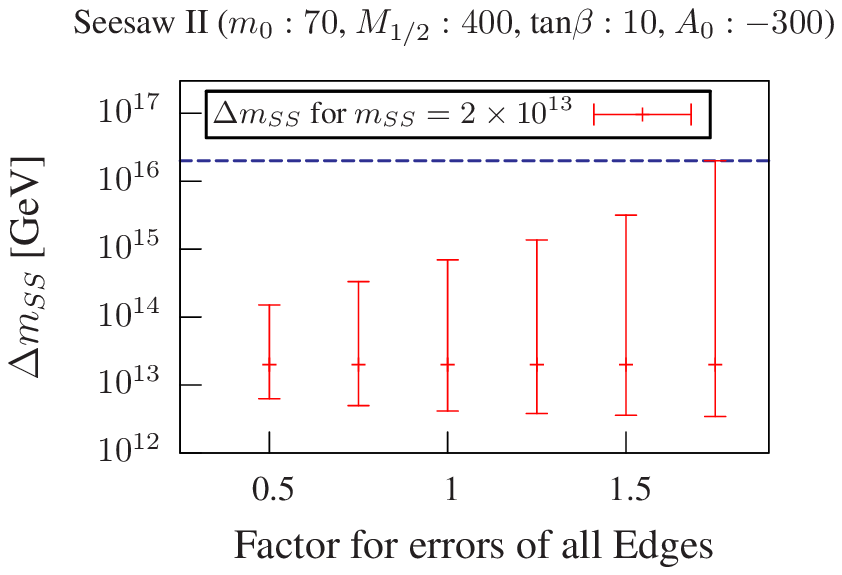}}
  \end{tabular}
  \caption{\label{fig:error_sbottomEdge}Error of $m_{SS}$ with respect
  to the error of $(m_{\tilde{g}} - m_{\tilde{b}_1})$ and
  $(m_{ll})^{\text{edge}}$. According to \cite{Weiglein:2004hn} the 
  error on $(m_{\tilde{g}} - m_{\tilde{b}_1})$ is expected to be 
  $\sim3.5\%$ and the error of $(m_{ll})^{\text{edge}}$ is estimated
  as $\sim0.17\%$. Note that we also changed the error of 
  $(m_{\tilde{g}} - m_{\tilde{b}_2})$, accordingly. }
 \end{center}
\end{figure}

We now turn to the discussion of LHC errors. In this case we do not 
use any input from the ILC. Fig. (\ref{fig:error_sbottomEdge}) shows 
$\Delta(m_{SS})$ as a function of the assumed error. 
Two observables are shown:
to the left as a function of $\Delta_{{\tilde g}{\tilde b}_i}$ and to 
the right as a function of the edge variables. Note that 
for SPS1a the LHC error for $\Delta_{{\tilde g}{\tilde b}_1}$ is estimated 
to be $\sim 3.5$ \%, while $\Delta(m_{ll}^{\text{edge}})$ should be 
measured to an accuracy of $0.17$ \%. Note that, while 
$\Delta(m_{ll}^{\text{edge}})$ can possibly be accurately measured 
in wide ranges of mSugra parameter space, the accuracy with which 
$\Delta_{{\tilde g}{\tilde b}_i}$ can be measured is far less certain. 
Both smaller and much larger errors on this quantity 
have been found in different study points, see \cite{Weiglein:2004hn}. 

The figure shows that for $m_{SS}=2\times 10^{13}$ a $7$ \% error 
on $\Delta_{{\tilde g}{\tilde b}_1}$ (which corresponds to a factor 
of 2 in the plot) is sufficient to distinguish 
between pmSugra and type-II, while an error of $9$ \% on this quantity 
is not sufficient. Again, the maximum value of this error which still 
allows to distinguish between type-II and pmSugra is a strong 
function of the (unknown) $m_{SS}$ itself. However, we have found 
that always $\Delta_{{\tilde g}{\tilde b}_1}$ is a critical input 
observable for our analysis. \footnote{We also consider 
$\Delta_{{\tilde g}{\tilde b}_2}$, which, however, is less important 
due to its larger error.} The importance of $\Delta_{{\tilde g}{\tilde b}_1}$ 
can be understood from Fig. (\ref{fig:running_masses_2}) and 
(\ref{fig:running_LHCobervables}): coloured sparticle 
masses depend much more strongly on $m_{SS}$ than masses of, for 
example, sleptons. Thus, despite the larger relative error on 
$\Delta_{{\tilde g}{\tilde b}_1}$ compared to the edge variables,it 
is nearly as important as demonstrated in fig. (\ref{fig:error_sbottomEdge}) 
to the right.

\begin{figure}
 \begin{center}
  \resizebox{80mm}{!}{\includegraphics
   {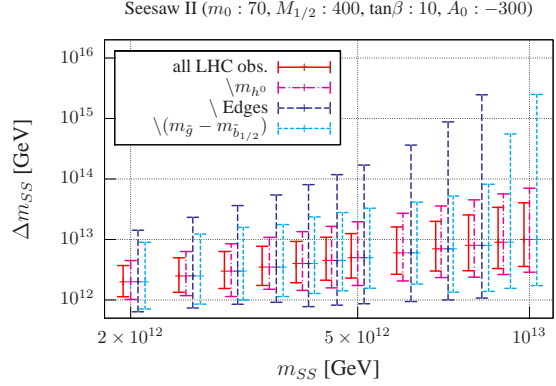}}
  \caption{\label{fig:error_LHC_SPS1App_ObsONOFF_SII} In this plot the
  errors of $m_{SS}$ with respect to $m_{SS}$ are shown. The different
  lines belong to different runs at which we switched off different
  combinations of LHC observables. The different lines belong in each
  case to the same $m_{SS}$ values but we separated them in the plot a
  bit to be able to distinguish them. The real $m_{SS}$ value is the
  value of the pink error bars that correspond to the run in which 
  the Higgs mass measurement is disabled.}
 \end{center}
\end{figure}

Fig.\ (\ref{fig:error_LHC_SPS1App_ObsONOFF_SII})  shows the 
results of different runs, where we have switched off artificially 
different combinations of observables. As noted above, 
$\Delta_{{\tilde g}{\tilde b}_1}$ and the edges are the most important 
observables for fixing $\Delta(m_{SS})$.  However, the Higgs mass 
measurement is not negligible, despite the fact that $\Delta(m_{SS})$ 
does not increase much in the figure, when $m_{h^0}$ is switched 
off. This importance lies in the fact that without $m_{h^0}$
the largest value of $m_{SS}$ not compatible with $M_G$ is 
$2 \times 10^{13}$ GeV, compared to $m_{SS}= 10^{14}$ GeV for
$m_{h^0}$ switched on.

We do not repeat the discussion for type-III seesaw. Results are 
very similar qualitatively, but again larger values of $m_{SS}$ 
can be tested for the same errors on the observables. 

\begin{figure}
 \begin{center}
  \resizebox{80mm}{!}{\includegraphics
   {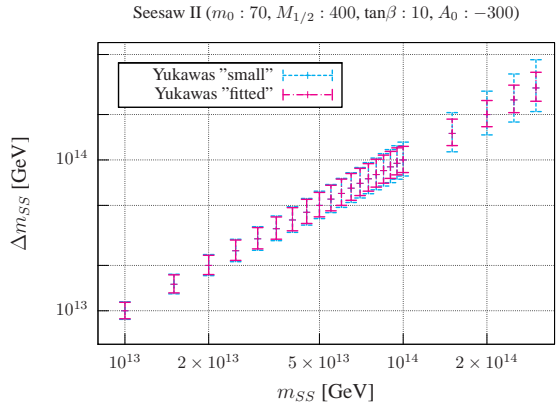}}
  \caption{\label{fig:error_ALL_yukawas_SPS1App_SII} In this plot the
  errors of $m_{SS}$ with respect to $m_{SS}$ are shown. The different
  lines belong to two different runs at which we set the Yukawa
  couplings to very small values ($\sim10^{-4}$) in the first run and
  in the second one we fitted the Yukawa couplings to neutrino
  data and calculated the error with the fitted values.}
 \end{center}
\end{figure}

Finally, we turn to the question of Yukawa couplings. Fig.\ 
(\ref{fig:error_ALL_yukawas_SPS1App_SII}) 
shows again $\Delta(m_{SS})$ 
as a function of $m_{SS}$ for two 
different calculations: (i)  a calculation with triplet Yukawa 
couplings negligibly small (all $(Y_T)_{ij}\sim {\cal O}(10^{-4})$) 
and (ii)  a calculation in which $Y_T$ has been fitted to give the 
atmospheric and solar neutrino mass squareds at their best 
fit values with neutrino angles taking tri-bimaximal values. 
As can be seen, differences between both calculations become 
negligible below roughly $m_{SS}=10^{14}$ GeV, as expected. 
For larger values of $m_{SS}$ correctly fitting the Yukawas leads 
to slightly smaller errors on $m_{SS}$. This can be understood, 
since for finite Yukawas SUSY masses change slightly stronger than 
for infinitesimal values of $Y_T$, making the fit easier. Note, 
however, that we have not scanned over all allowed range of 
$Y_T$ in this calculation.  In a complete $14$ parameter
$\chi^2$ fit errors might be larger. Note also, we can not find 
any good neutrino solution for $m_{SS}$ larger than $m_{SS}=6\times 10^{14}$, 
since in this calculation we have chosen for the coupling $\lambda_2=0.5$. 
In conclusion, a full fit including Yukawa couplings will be 
necessary only if signs of $m_{SS}\gsim 10^{14}$ GeV have been 
found in SUSY mass data.

\section{Conclusions and discussion}

We have studied the possibility to obtain indirect information on 
the seesaw scale from SUSY mass measurements at future colliders. 
Since in the type-I seesaw only SM singlets are added to the MSSM 
particle content, none of the measurements which we considered are 
expected to show sizeable departures from mSugra expectations. We 
therefore concentrated on the type-II and type-III realizations 
of the seesaw. 

Assuming mSugra boundary conditions and taking error estimates as 
forecasted by study groups we find that a combination of LHC and 
ILC measurements should be able to distinguish pure mSugra from 
mSugra plus either type-II or type-III seesaw for nearly any 
relevant values of the seesaw scale, if (a) at least $\chi^0_1$, 
${\tilde e}_R$/ ${\tilde \mu}_R$ and ${\tilde e}_L$/ ${\tilde \mu}_L$ 
are kinemetically accessible at the ILC and (b) the LHC can measure 
$m_{\tilde g}-m_{{\tilde b}_1}$ and the edge observables accurately. 
We always assume that the lightest Higgs has been found. 
The degree of confidence with which pmSugra can be 
distinguished from mSugra plus seesaw depends sensitively on the 
actual value of the seesaw scale. At the ``critical'' value of 
$m_{SS} \sim 10^{15}$, beyond which neutrino data can no longer be 
explained with Yukawas smaller than 1, the difference between 
type-II + mSugra and pmSugra could be as low as only 1 $\sigma$ c.l. 
However, the difference between pmSugra and mSugra + seesaw rises 
very sharply with decreasing $m_{SS}$ and formally more than 5 $\sigma$ 
c.l. could be reached already at $(5-6) \times 10^{14}$ for type-II. 
Differences between pmSugra and mSugra plus type-III are always found 
to be larger than for mSugra plus type-II for the same value of 
$m_{SS}$.

As expected, the future is not as bright, if we take into account 
only LHC data. Nevertheless, using only LHC data one
can distinguish pure mSugra and 
mSugra plus seesaw in some favourable parts of 
parameter space. Especially, we point out that the lower the real 
value of the seesaw scale $m_{SS}$ is, the easier it becomes to 
distinguish pure mSugra from mSugra plus seesaw. We have discussed 
the most important measurements for the LHC and the ILC and the 
relative errors with which these observables need to be measured 
for this analysis to be possible.
In our analysis we used exclusively mSugra SUSY breaking boundary 
conditions, but other, more involved SUSY breaking schemes with 
more free parameters could, in principle, be ``tested'' in a 
similar way. 

Of course, the analysis presented in this paper is far from being 
complete. If SUSY is found at the LHC, one would need to redo our 
calculations with real data. However, the experimentalist would not, 
as we have always assumed in our fits, know the real values of the 
parameters. We have tried for a few points, whether the correct input 
parameters can be retrieved for arbitrary starting points in our MC 
random walk procedure and - given enough CPU-time - are able to find 
the correct minimum. However, our ``observables'' are theoretically 
calculated observables and thus perfect in contrast to real data which 
are expected to scatter around the true values and might show tension 
between different observables.  Thus, finding the correct minimum in real 
data might be more difficult. Moreover, the underlying model
will not be known a priori and, thus,  $\chi^2_{Min}$ for different 
models need to be calculated and compared.

Nevertheless, even taking into account the limitations of our study, 
we think it is highly motivating that type-II and type-III seesaw 
leave sizeable traces in SUSY spectra, which should show up, 
if sufficiently accurate mass measurements 
are possible and become available.

\section*{Acknowledgments}

W.P. thanks IFIC/C.S.I.C. for hospitality during an extended stay.
This work was supported by the Spanish MICINN under grants
FPA2008-00319/FPA, by the MULTIDARK Consolider CSD2009-00064, by
Prometeo/2009/091, by the EU grant UNILHC PITN-GA-2009-237920. 
W.P. is  supported by the DFG, project number PO-1337/1-1, 
 and by the Alexander 
von Humboldt Foundation.


\begin{thebibliography}{10}

\bibitem{Chamseddine:1982jx}
  A.~H.~Chamseddine, R.~L.~Arnowitt and P.~Nath,
  Phys.\ Rev.\ Lett.\  {\bf 49} (1982) 970.

\bibitem{Nilles:1983ge}
   H.~P.~Nilles,
  Phys.\ Rept.\  {\bf 110} (1984) 1.


\bibitem{Giudice:1998xp}
  G.~F.~Giudice, M.~A.~Luty, H.~Murayama and R.~Rattazzi,
  JHEP {\bf 9812} (1998) 027
  [arXiv:hep-ph/9810442].

\bibitem{Chacko:1999am}
  Z.~Chacko, M.~A.~Luty, I.~Maksymyk and E.~Ponton,
  JHEP {\bf 0004} (2000) 001
  [arXiv:hep-ph/9905390].

\bibitem{Giudice:1998bp}
  For a review on GMSB, see: 
  G.~F.~Giudice and R.~Rattazzi,
  Phys.\ Rept.\  {\bf 322} (1999) 419
  [arXiv:hep-ph/9801271].


\bibitem{AguilarSaavedra:2001rg}
  J.~A.~Aguilar-Saavedra {\it et al.}  [ECFA/DESY LC Physics Working Group],
  arXiv:hep-ph/0106315.

\bibitem{Weiglein:2004hn}
  G.~Weiglein {\it et al.}  [LHC/LC Study Group],
  Phys.\ Rept.\  {\bf 426} (2006) 47
  [arXiv:hep-ph/0410364].

\bibitem{AguilarSaavedra:2005pw}
  J.~A.~Aguilar-Saavedra {\it et al.},
  Eur.\ Phys.\ J.\  C {\bf 46}, 43 (2006)
  [arXiv:hep-ph/0511344].

\bibitem{Blair:2000gy}
  G.~A.~Blair, W.~Porod and P.~M.~Zerwas,
  Phys.\ Rev.\  D {\bf 63} (2001) 017703
  [arXiv:hep-ph/0007107].

\bibitem{Blair:2002pg}
G.~A. Blair, W.~Porod and P.~M. Zerwas,
\newblock Eur. Phys. J. {\bf C27}, 263 (2003), [hep-ph/0210058].

\bibitem{Bechtle:2005vt}
  P.~Bechtle, K.~Desch, W.~Porod and P.~Wienemann,
  Eur.\ Phys.\ J.\  C {\bf 46} (2006) 533
  [arXiv:hep-ph/0511006].

\bibitem{Lafaye:2007vs}
  R.~Lafaye, T.~Plehn, M.~Rauch and D.~Zerwas,
  Eur.\ Phys.\ J.\  C {\bf 54} (2008) 617
  [arXiv:0709.3985 [hep-ph]].

\bibitem{Adam:2010uz}
  C.~Adam, J.~L.~Kneur, R.~Lafaye, T.~Plehn, M.~Rauch and D.~Zerwas,
  arXiv:1007.2190 [hep-ph].

\bibitem{Fukuda:1998mi}
Y.~Fukuda {\it et al.}  [Super-Kamiokande Collaboration],
Phys.\ Rev.\ Lett.\  {\bf 81}, 1562 (1998)
\bibitem{Ahmad:2002jz}
SNO, Q.~R. Ahmad {\em et~al.},
\newblock Phys. Rev. Lett. {\bf 89}, 011301 (2002), [nucl-ex/0204008].
\bibitem{Eguchi:2002dm}
KamLAND, K.~Eguchi {\em et~al.},
\newblock Phys. Rev. Lett. {\bf 90}, 021802 (2003), [hep-ex/0212021].

\bibitem{KamLAND2007}
 KamLAND~Collaboration,
 arXiv:0801.4589 [hep-ex].


\bibitem{Schwetz:2008er}
  For a recent review on the status of neutrino oscillation data, see: 
  T.~Schwetz, M.~A.~Tortola and J.~W.~F.~Valle,
  New J.\ Phys.\  {\bf 10}, 113011 (2008)
  [arXiv:0808.2016 [hep-ph]]. 
  Version 3 on the arXive is updated with data until Feb 2010 


\bibitem{Minkowski:1977sc}
  P.~Minkowski,
  Phys.\ Lett.\ B {\bf 67} (1977) 421.

\bibitem{seesaw}
T.~Yanagida, in {\it KEK lectures}, ed.  O.~Sawada and A.~Sugamoto,
KEK, 1979;
M Gell-Mann, P Ramond, R. Slansky, in {\it Supergravity}, ed. P. van
Niewenhuizen and D. Freedman (North Holland, 1979);

\bibitem{MohSen}
R.N.~Mohapatra and G.~Senjanovic, {\sl Phys. Rev. Lett.}\/ {\bf 44}
912 (1980).

\bibitem{Ma:1998dn}
  E.~Ma,
  Phys.\ Rev.\ Lett.\  {\bf 81}, 1171 (1998)
  [arXiv:hep-ph/9805219].

\bibitem{Schechter:1980gr}
  J.~Schechter and J.~W.~F.~Valle,
  Phys.\ Rev.\ D {\bf 22}, 2227 (1980).

\bibitem{Cheng:1980qt}
  T.~P.~Cheng and L.~F.~Li,
  Phys.\ Rev.\  D {\bf 22}, 2860 (1980).

\bibitem{Foot:1988aq}
  R.~Foot, H.~Lew, X.~G.~He and G.~C.~Joshi,
  Z.\ Phys.\  C {\bf 44}, 441 (1989).

\bibitem{Hisano:1995cp}
A rather incomplete list on LFV in SUSY seesaw, mainly on type-I is:
J.~Hisano, T.~Moroi, K.~Tobe and M.~Yamaguchi,
\newblock Phys. Rev. {\bf D53}, 2442 (1996);
J.~R.~Ellis, J.~Hisano, M.~Raidal and Y.~Shimizu,
Phys.\ Rev.\  D {\bf 66}, 115013 (2002);
F.~Deppisch, H.~Paes, A.~Redelbach, R.~R\"uckl and Y.~Shimizu,
\newblock Eur. Phys. J. {\bf C28}, 365 (2003);
S.~T.~Petcov, S.~Profumo, Y.~Takanishi and C.~E.~Yaguna,
Nucl.\ Phys.\  B {\bf 676} (2004) 453;
E.~Arganda and M.~J. Herrero,
\newblock Phys. Rev. {\bf D73}, 055003 (2006);
S.~T.~Petcov, T.~Shindou and Y.~Takanishi,
Nucl.\ Phys.\  B {\bf 738}, 219 (2006);
S.~Antusch, E.~Arganda, M.~J. Herrero and A.~M. Teixeira,
\newblock JHEP {\bf 11}, 090 (2006);
F.~Deppisch and J.~W.~F.~Valle,
Phys.\ Rev.\  D {\bf 72}, 036001 (2005);
J.~Hisano, T.~Moroi, K.~Tobe, M.~Yamaguchi and T.~Yanagida,
\newblock Phys. Lett. {\bf B357}, 579 (1995);
E.~Arganda, M.~J. Herrero and A.~M. Teixeira,
\newblock JHEP {\bf 10}, 104 (2007), [0707.2955];
F.~Deppisch, T.~S. Kosmas and J.~W.~F. Valle,
\newblock Nucl. Phys. {\bf B752}, 80 (2006), [hep-ph/0512360].

\bibitem{Borzumati:1986qx}
  F.~Borzumati and A.~Masiero,
  Phys.\ Rev.\ Lett.\  {\bf 57}, 961 (1986).

\bibitem{Rossi:2002zb}
  A.~Rossi,
  Phys.\ Rev.\  D {\bf 66}, 075003 (2002)

\bibitem{Hirsch:2008gh}
  M.~Hirsch, S.~Kaneko and W.~Porod,
  Phys.\ Rev.\  D {\bf 78}, 093004 (2008)


\bibitem{Joaquim:2006uz}
  F.~R.~Joaquim and A.~Rossi,
  Phys.\ Rev.\ Lett.\  {\bf 97} (2006) 181801
  [arXiv:hep-ph/0604083].

\bibitem{Joaquim:2006mn}
  F.~R.~Joaquim and A.~Rossi,
  Nucl.\ Phys.\  B {\bf 765} (2007) 71
  [arXiv:hep-ph/0607298].

\bibitem{Brignole:2010nh}
  A.~Brignole, F.~R.~Joaquim and A.~Rossi,
  JHEP {\bf 1008} (2010) 133
  [arXiv:1007.1942 [hep-ph]].


\bibitem{Hirsch:2008dy}
M.~Hirsch, J.~W.~F.~Valle, W.~Porod, J.~C.~Romao and A.~Villanova del Moral,
Phys.\ Rev.\  D {\bf 78}, 013006 (2008)


\bibitem{Davidson:2001zk}
  S.~Davidson and A.~Ibarra,
  JHEP {\bf 0109} (2001) 013
  [arXiv:hep-ph/0104076].

\bibitem{Ibarra:2005qi}
  A.~Ibarra,
  JHEP {\bf 0601} (2006) 064
  [arXiv:hep-ph/0511136].



\bibitem{Freitas:2005et}
A.~Freitas, W.~Porod and P.~M. Zerwas,
\newblock Phys. Rev. {\bf D72}, 115002 (2005), [hep-ph/0509056].

\bibitem{Deppisch:2007xu}
  F.~Deppisch, A.~Freitas, W.~Porod and P.~M.~Zerwas,
  Phys.\ Rev.\  D {\bf 77} (2008) 075009
  [arXiv:0712.0361 [hep-ph]].

\bibitem{Kadota:2009sf}
  K.~Kadota and J.~Shao,
  Phys.\ Rev.\  D {\bf 80} (2009) 115004
  [arXiv:0910.5517 [hep-ph]].

%
\bibitem{Allanach:2008ib}
  B.~C.~Allanach, J.~P.~Conlon and C.~G.~Lester,
  Phys.\ Rev.\  D {\bf 77}, 076006 (2008)
  [arXiv:0801.3666 [hep-ph]].

\bibitem{Buras:2009sg}
  A.~J.~Buras, L.~Calibbi and P.~Paradisi,
  JHEP {\bf 1006} (2010) 042
  [arXiv:0912.1309 [hep-ph]].

\bibitem{Abada:2010kj}
  A.~Abada, A.~J.~R.~Figueiredo, J.~C.~Romao and A.~M.~Teixeira,
  JHEP {\bf 1010}, 104 (2010)
  [arXiv:1007.4833 [hep-ph]].



\bibitem{Calibbi:2007bk}
  L.~Calibbi, Y.~Mambrini and S.~K.~Vempati,
  JHEP {\bf 0709}, 081 (2007)
  [arXiv:0704.3518 [hep-ph]].

\bibitem{Kadota:2009vq}
  K.~Kadota, K.~A.~Olive and L.~Velasco-Sevilla,
  Phys.\ Rev.\  D {\bf 79}, 055018 (2009)
  [arXiv:0902.2510 [hep-ph]].

\bibitem{Kadota:2009fg}
  K.~Kadota and K.~A.~Olive,
  Phys.\ Rev.\  D {\bf 80}, 095015 (2009)
  [arXiv:0909.3075 [hep-ph]].

\bibitem{Kang:2010zd}
  S.~K.~Kang, T.~Morozumi and N.~Yokozaki,
  JHEP {\bf 1011} (2010) 061
  [arXiv:1005.1354 [hep-ph]].

\bibitem{Heinemeyer:2010eg}
  S.~Heinemeyer, M.~J.~Herrero, S.~Penaranda and A.~M.~Rodriguez-Sanchez,
  arXiv:1007.5512 [hep-ph].

\bibitem{Esteves:2009qr}
  J.~N.~Esteves, S.~Kaneko, J.~C.~Romao, M.~Hirsch and W.~Porod,
  Phys.\ Rev.\  D {\bf 80}, 095003 (2009)
  [arXiv:0907.5090 [hep-ph]].

\bibitem{Esteves:2010ff}
  J.~N.~Esteves, J.~C.~Romao, M.~Hirsch, F.~Staub and W.~Porod,
  Phys.\ Rev.\  D {\bf 83} (2011) 013003
  [arXiv:1010.6000 [hep-ph]].


\bibitem{Buckley:2006nv}
M.~R.~Buckley and H.~Murayama,
Phys.\ Rev.\ Lett.\  {\bf 97}, 231801 (2006)
[arXiv:hep-ph/0606088].




\bibitem{Borzumati:2009hu}
 F.~Borzumati and T.~Yamashita,
  Prog.\ Theor.\ Phys.\  {\bf 124} (2010) 761
  [arXiv:0903.2793 [hep-ph]].




\bibitem{Porod:2003um}
  W.~Porod,
  Comput.\ Phys.\ Commun.\  {\bf 153}, 275 (2003)
  [arXiv:hep-ph/0301101].

\bibitem{SPheno}
For the latetst version of SPheno, see the web page: 
http://www.physik.uni-wuerzburg.de/$\sim$porod/SPheno.html

\bibitem{Staub:2008uz}
  F.~Staub,
  arXiv:0806.0538 [hep-ph].

\bibitem{Staub:2009bi}
  F.~Staub,
  Comput.\ Phys.\ Commun.\  {\bf 181}, 1077 (2010)
  [arXiv:0909.2863 [hep-ph]].


\bibitem{Staub:2010jh}
  F.~Staub,
  arXiv:1002.0840 [hep-ph].

\bibitem{Allanach:2002nj}
B.~C.~Allanach {\it et al.},
in {\it Proc. of the APS/DPF/DPB Summer Study on the Future of 
Particle Physics (Snowmass 2001) } ed. N.~Graf,
Eur.\ Phys.\ J.\  C {\bf 25}, 113 (2002)
[arXiv:hep-ph/0202233].

\bibitem{Bachacou:1999zb}
  H.~Bachacou, I.~Hinchliffe and F.~E.~Paige,
  Phys.\ Rev.\  D {\bf 62} (2000) 015009
  [arXiv:hep-ph/9907518].

\bibitem{Allanach:2000kt}
  B.~C.~Allanach, C.~G.~Lester, M.~A.~Parker and B.~R.~Webber,
  JHEP {\bf 0009} (2000) 004
  [arXiv:hep-ph/0007009].

\bibitem{Lester:2001zx}
  C.~G.~Lester,
  ``Model independent sparticle mass measurements at ATLAS''; 
  CERN-THESIS-2004-003

\bibitem{Pierce:1996zz}
  D.~M.~Pierce, J.~A.~Bagger, K.~T.~Matchev and R.~j.~Zhang,
  Nucl.\ Phys.\  B {\bf 491} (1997) 3
  [arXiv:hep-ph/9606211].

\bibitem{Nakamura:2010zzi}
  K.~Nakamura {\it et al.}  [Particle Data Group],
  J.\ Phys.\ G {\bf 37} (2010) 075021.

\bibitem{LEP2online}
http://lepsusy.web.cern.ch/lepsusy/


\bibitem{TevatronBounds} J.~Yamaoka, talk given at PASCOS 2010, Valencia
  (Spain), July 19th - 23rd, 2010.

\end{thebibliography}
\end{document}